%% file: fplane_prep_proof.tex
\documentclass[twocolumn]{aastex61}
\usepackage{xcolor}

\received{July 1, 20xx}
\revised{September xx, 20xx}
\accepted{\today}
\shorttitle{Discovery of a fundamental plane}
\shortauthors{Fujita et al.}

\begin{document}

\title{Discovery of a new fundamental plane dictating galaxy cluster
evolution from gravitational lensing}

\correspondingauthor{Yutaka Fujita}
\email{fujita@vega.ess.sci.osaka-u.ac.jp}

\author[0000-0003-0058-9719]{Yutaka Fujita}
\affil{Department of Earth and Space Science, Graduate School
 of Science, Osaka University, Toyonaka, Osaka 560-0043, Japan}

\author{Keiichi Umetsu}
\affiliation{Institute of Astronomy and Astrophysics, Academia Sinica,
 P.O. Box 23-141, Taipei 10617, Taiwan}

\author{Elena Rasia}
\affiliation{INAF, Osservatorio Astronomico di Trieste, via Tiepolo 11,
 I-34131, Trieste, Italy}

\author{Massimo Meneghetti} 
\affiliation{INAF, Osservatorio Astronomico
di Bologna, via Ranzani 1, I-40127 Bologna, Italy}

\author{Megan Donahue}
\affiliation{Physics and Astronomy Department, 
Michigan State University, East Lansing, MI, 48824 USA}

\author{Elinor Medezinski}
\affiliation{Department of Astrophysical Sciences, 
4 Ivy Lane, Princeton, NJ  08544, USA}

\author{Nobuhiro Okabe}
\affiliation{Department of Physical Science, Hiroshima University, 1-3-1
 Kagamiyama, Higashi-Hiroshima, Hiroshima 739-8526, Japan}

\author{Marc Postman}
\affiliation{Space Telescope Science Institute, 3700 San Martin Drive,
 Baltimore, MD 21208, USA}

\begin{abstract}

In cold dark-matter (CDM) cosmology, objects in the universe have grown
under the effect of gravity of dark matter. The intracluster gas in a
galaxy cluster was heated when the dark-matter halo formed through
gravitational collapse. The potential energy of the gas was converted to
thermal energy through this process. However, this process and the
thermodynamic history of the gas have not been clearly characterized in
connection with with the formation and evolution of the internal
structure of dark-matter halos. Here, we show that observational CLASH
data of high-mass galaxy clusters lie on a plane in the
three-dimensional logarithmic space of their characteristic radius
$r_s$, mass $M_s$, and X-ray temperature $T_X$ with a very small
orthogonal scatter. The tight correlation indicates that the gas
temperature was determined at a specific cluster formation time, which
is encoded in $r_s$ and $M_s$. The plane is tilted with respect to $T_X
\propto M_s/r_s$, which is the plane expected in the case of simplified
virial equilibrium. We show that this tilt can be explained by a
similarity solution, which indicates that clusters are not isolated but
continuously growing through matter accretion from their outer
environments. Numerical simulations reproduce the observed plane and its
angle. This result holds independently of the gas physics implemented in
the code, revealing the fundamental origin of this plane.

\end{abstract}

keywords{galaxies: clusters: general --- cosmology: theory --- dark
matter --- large-scale structure of Universe}

\section{Introduction} 
\label{sec:intro}

$N$-body numerical simulations show that the density profile of
dark-matter halos in galaxy clusters can be well described by the
Navarro--Frenk--White (\citealp{nav97a}; NFW, hereafter) density
profile: $\rho_{\rm DM}\propto (r/r_s)^{-1}(1+r/r_s)^{-2}$, where $r$ is
the clustercentric distance and $r_s$ is the characteristic or scale
radius. We define the mass inside $r_s$ as $M_s$. Recent
higher-resolution simulations have shown that the internal structure of
dark-matter halos reflects the growth history of the halos
\citep{wec02a,zha03a,lud13a,cor15b,mor15b}. They show that in the early
``fast-rate growth'' phase, halos grow rapidly through massive matter
accumulation. This growth is often associated with phenomena that erase
the previous internal structure of the halos, such as major mergers with
other halos. In the subsequent “slow-rate growth” phase, halos
gradually grow through moderate matter accretion from their
surroundings. There are multiple definitions of the formation time of a
halo. One example is the time at which the mass of the main progenitor
equals the characteristic mass $M_s$ of the $z=0$ halo
\citep{lud13a,cor15b}, and it approximately represents the end of the
fast-growth phase and the transitioning toward the slow-growth phase.
Only the outskirts of the halos ($r> r_s$) gradually grow in the latter
phase \citep{wec02a,zha03a,lud13a,cor15b,mor15b}. Thus, dark-matter
halos are assembled from the inside out, and the results of the
numerical simulations can be interpreted such that the characteristic
radius $r_s$ and the mass $M_s$ preserve a memory of the formation time
of the halo \citep{wec02a,zha03a,lud13a,cor15b}.  In this ``inside-out''
scenario of halo formation, halos take a range of characteristic
densities ($\rho_s\equiv 3 M_s/(4\pi r_s^3)$); older halos tend to be
more concentrated and have larger characteristic densities, which
reflects the higher average density of the universe in the past
\citep{nav97a,wec02a,zha03a,lud13a,cor15b,mor15b}. This scenario is in
contrast with the classical approach in which halos are continuously
modified, even by minor mergers, and constantly changing their profiles
so that dark-matter halos lose the memory of their epoch of formation
\citep{gun72a,1974ApJ...187..425P}.

If the inside-out halo growth scenario is correct, then we would expect
that the formation time not only reflects the structural parameters
($r_s$ and $M_s$) in the form of the characteristic density ($\rho_s$),
but also influences the properties of the X-ray intracluster
gas. However, its quantitative dependence is not obvious because the gas
is collisional matter in contrast with dark matter. The hot gas is
expected to be heated mostly via merger shocks produced when smaller
halos fall into the halo \citep{ras11a,kra12a}. However, it is difficult
to directly observe the heating process, as the shocks are often located
at the outskirts of clusters \citep{2000ApJ...542..608M,ryu03a}, where
the gas emission is very faint. We here investigate correlations between
the halo parameters ($r_s, M_s$) and the average X-ray gas temperature
$T_X$ because the temperature is supposedly sensitive to the depth of
the potential well and the past heating process
\citep{1998ApJ...503..569E}. Since the emissivity of the X-ray gas is
proportional to the gas density squared, the average measured
temperature of a cluster mainly reflects the temperature in the region
($r\lesssim r_s$), where the density is high. While there were previous
studies that attempted to investigate correlations among a certain
combination of three cluster structural parameters
\citep{sch93a,ada98a,fuj99c,lan04b,2006ApJ...640..673O}, they adopted
parameters such as the galaxy luminosities that are not directly related
to the structure of halos.

In the paper, we assume a spatially flat $\Lambda$CDM cosmology with
$\Omega_\mathrm{m}=0.27$, $\Omega_\Lambda=0.73$, and the Hubble constant
of $H_0=70$\,km\,s$^{-1}$\,Mpc$^{-1}$ throughout this paper.

\section{Observational Data}

We study the Cluster Lensing And Supernova survey with Hubble (CLASH)
observational dataset that includes 20 massive clusters, most of which
are apparently relaxed, X-ray regular systems\footnote{Among the 20
clusters, 16 of them are X-ray selected and the rest are the CLASH
high-magnification clusters that may not be relaxed systems.}
\citep{pos12a,men14a}. The range of redshifts is 0.187--0.686, and
their median redshift is 0.377 \citep{ume16a}. Lensing constraints on
the NFW characteristic radius $r_s$ and $M_s$ were obtained from a joint
analysis of strong-lensing, weak-lensing shear and magnification data
of background galaxies \citep{ume16a}.  Their analysis is based on
16-band Hubble Space Telescope observations \citep{2015ApJ...801...44Z}
and wide-field multi-color imaging taken primarily with Suprime-Cam on
the Subaru Telescope \citep{2014ApJ...795..163U}. The core-excised X-ray
temperatures of the clusters were taken from \citet{pos12a}, in which
the temperatures are measured for the region of 50--500 kpc from the
cluster centers excluding the cool core at the center of the clusters.
We chose the outer radius considering the field of view of the {\it
Chandra} satellite and the completeness of the data. Since most of the
X-ray emissions come from $<500$~kpc, the increase of the radius does
not affect the results. The cluster data are shown in
Table~\ref{tab:data}.

\begin{figure*}
\plottwo{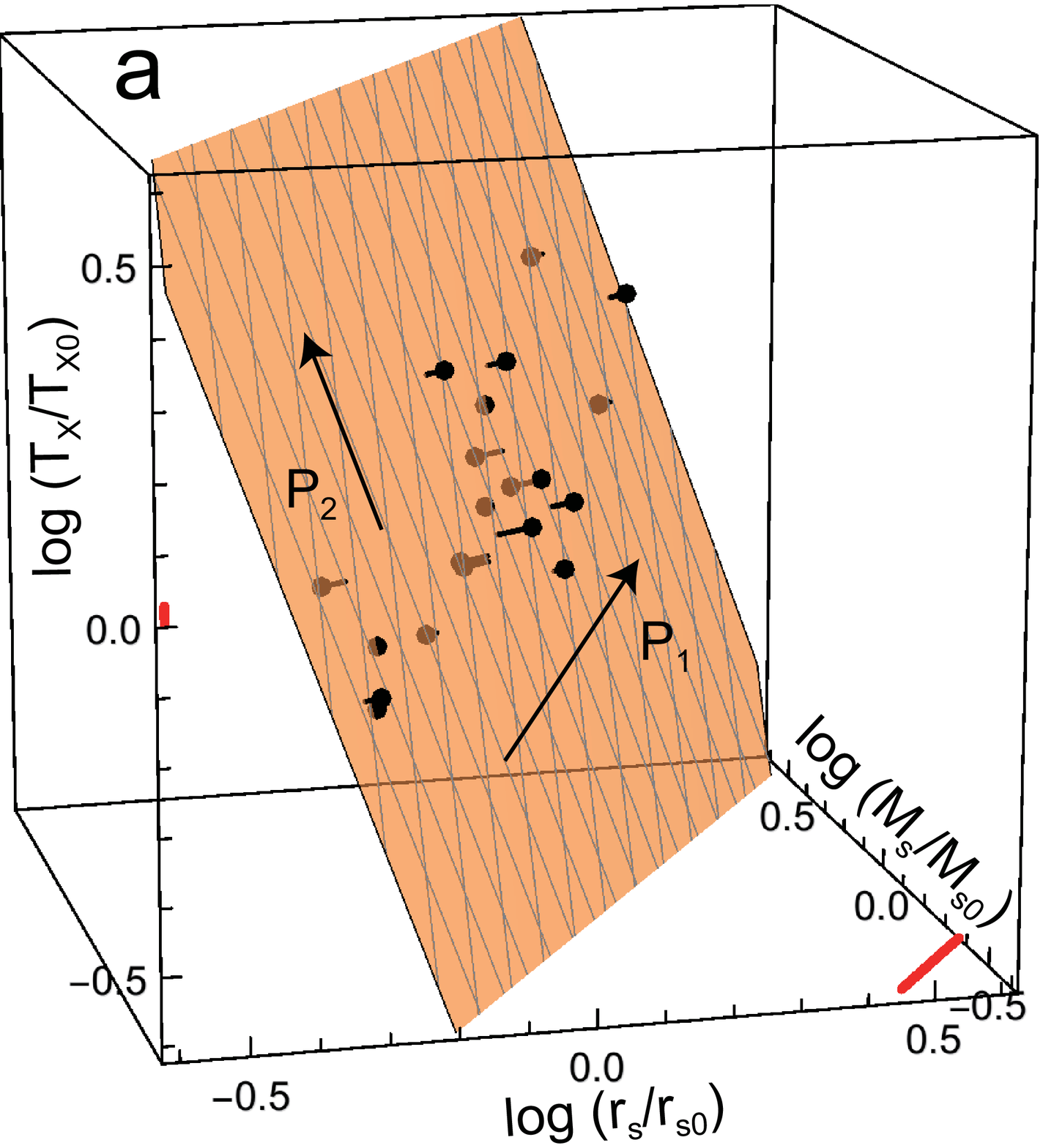}{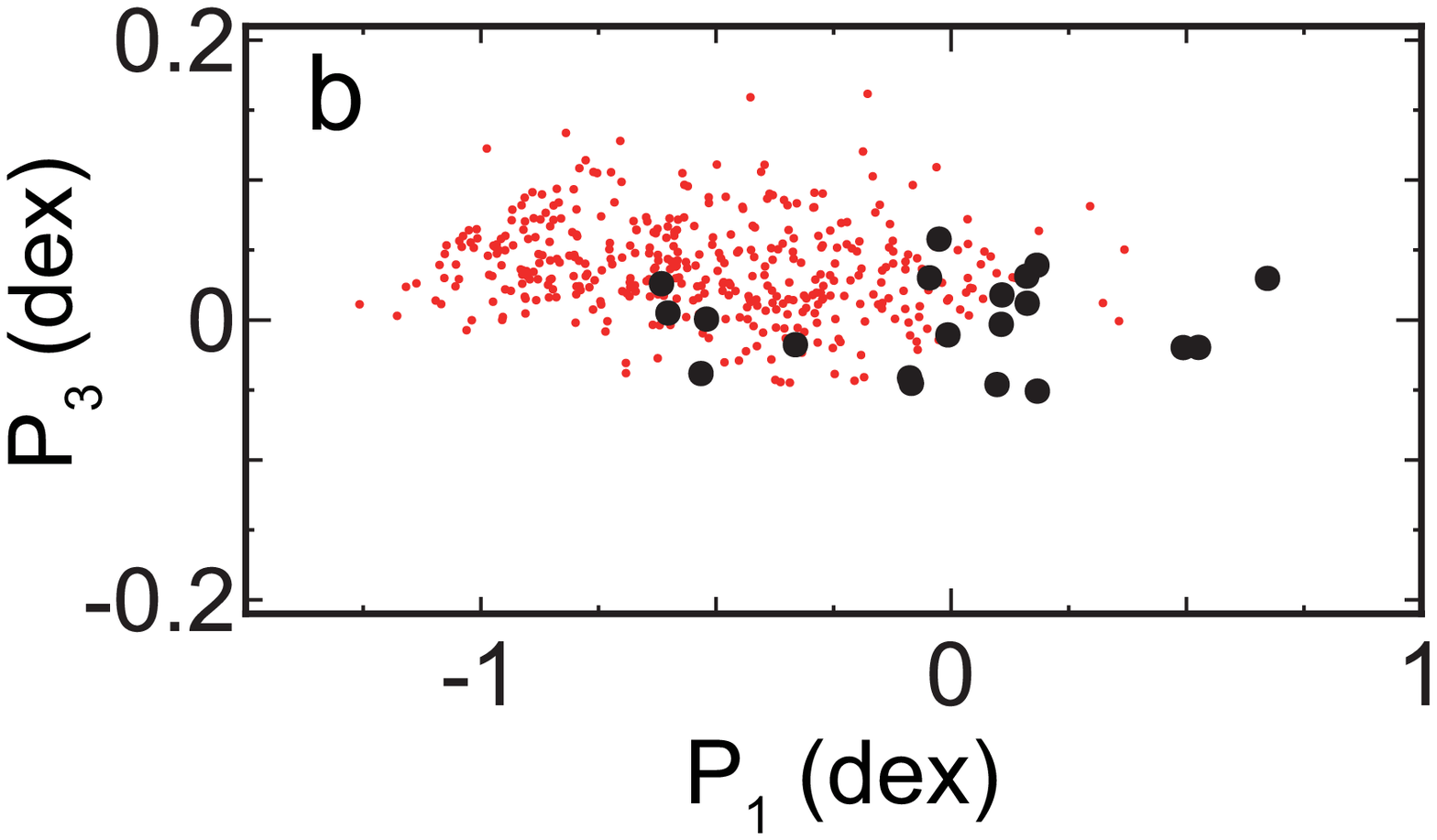} \epsscale{.50}\plotone{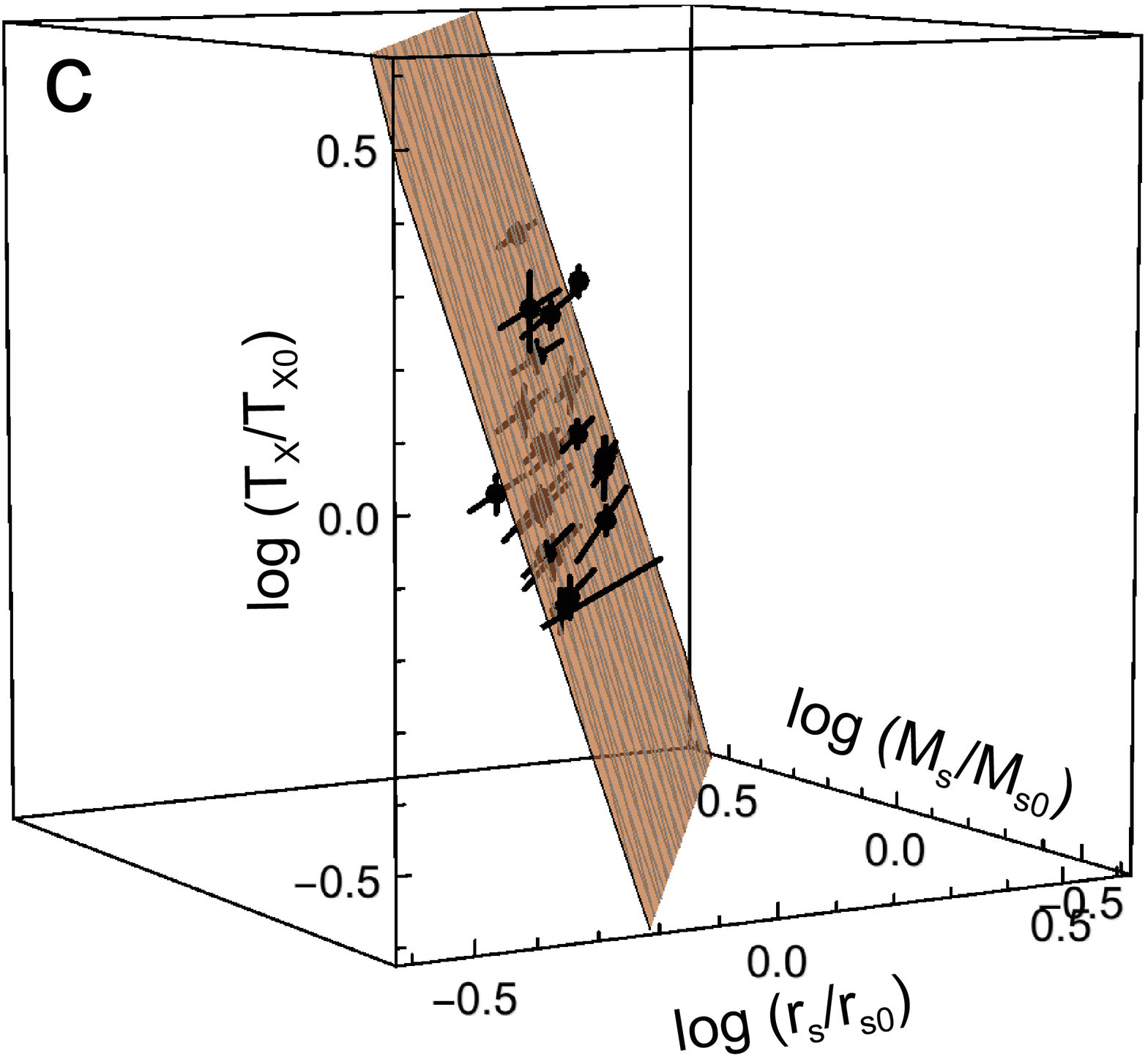}
\caption{(a) Points (pin heads) show the distribution of the observed
clusters in the space of $(\log (r_s/r_{s0}), \log (M_s/M_{s0}), \log
(T_{\rm X}/T_{\rm X0}))$, where $r_{s0}=570$~kpc, $M_{s0}=3.8\times
10^{14}\: M_\odot$, and $T_{\rm X0}=8.2$~keV are the sample geometric
averages (log means) of $r_s$, $M_s$, and $T_{\rm X}$, respectively. The
length of a pin shows the distance between the point and the obtained
plane. The orange plane is translucent and grayish points are located
below the plane. The arrow $P_1$ shows the direction on the plane in
which the data are most extended, and the arrow $P_2$ is perpendicular
to $P_1$ on the plane. Red bars at the corner of the $\log r_s$--$\log
M_s$ plane and on the $\log T_{\rm X}$ axis are typical $1\sigma$ errors
of the data. (b) The cross-section of the plane in (a). The origin is
the same as (a) and $P_3$ is the plane normal. Large black points are
the observations shown in (a). Small red points are the MUSIC simulated
clusters projected on the $P_1$--$P_3$ plane determined for the observed
clusters. (c) Same as (a) but error bars for individual clusters are
included. The viewing angle is changed so that the relation between the
error bars and the plane is easily seen. \label{fig:plane}}
\end{figure*}

\section{Fundamental Plane Analysis for the CLASH Sample}

In Figure~\ref{fig:plane}(a), we show the distribution of clusters in
the $(\log r_s, \log M_s, \log T_{\rm X})$ space. We see that the data
points are closely distributed on a plane that we have determined using
a principal component analysis (PCA) to minimize deviations of the data
points from the plane (see the \ref{sec:append}ppendix). The arrow $P_1$
shows the direction on the plane in which the data are most extended,
and the arrow $P_2$ is perpendicular to $P_1$ on the plane. The plane
normal is represented by $P_3$. The dispersion around the plane or the
thickness of the plane is shown in Figure~\ref{fig:plane}(b) and it is
only $0.045^{+0.008}_{-0.007}$ dex (all uncertainties are quoted at the
$1\:\sigma$ confidence level unless otherwise mentioned). The thickness
is comparable to that of the well-known fundamental plane for elliptical
galaxies in the space of the surface brightness, the effective radius,
and the velocity dispersion ($\sim 0.06$~dex; e.g.,
\citealp{lab08a,hyd09a}). In Figure~\ref{fig:plane}(c), we show error bars
for individual clusters. In the vertical direction ($T_{\rm X}$), we
show the temperature errors in Table~\ref{tab:data}. In the horizontal
direction, however, the errors of $r_s$ and $M_s$ are highly correlated,
and we display them as a single bar. That is, for each cluster, we draw
a bar connecting ($r_s^u$, $M_s^u$) and ($r_s^l$, $M_s^l$), where the
superscripts $u$ and $l$ are the upper and the lower limits shown in
Table~\ref{tab:data}, respectively.  Note that we have properly
accounted for the correlation for each cluster using the joint posterior
probability distribution of the NFW parameters (mass and concentration)
when we calculate the plane parameters (see the
\ref{sec:append}ppendix). Thus, the actual error is not represented by a
single bar in a precise sense.  The tight planar distribution in
Figure~\ref{fig:plane} indicates that the structure of the dark-matter
halos ($r_s$ and $M_s$) did make a direct influence on the properties of
the intracluster gas ($T_{\rm X}$).  In the context of the inside-out
halo growth scenario, the most natural interpretation of our findings is
that the intracluster gas was heated up to around $T_{\rm X}$ in the
fast-rate growth phase when the shape of the potential well ($r_s$ and
$M_s$) was established, and that the gas preserves the memory of the
cluster formation as is the case of the dark-matter halo structure.

\begin{figure}
\plotone{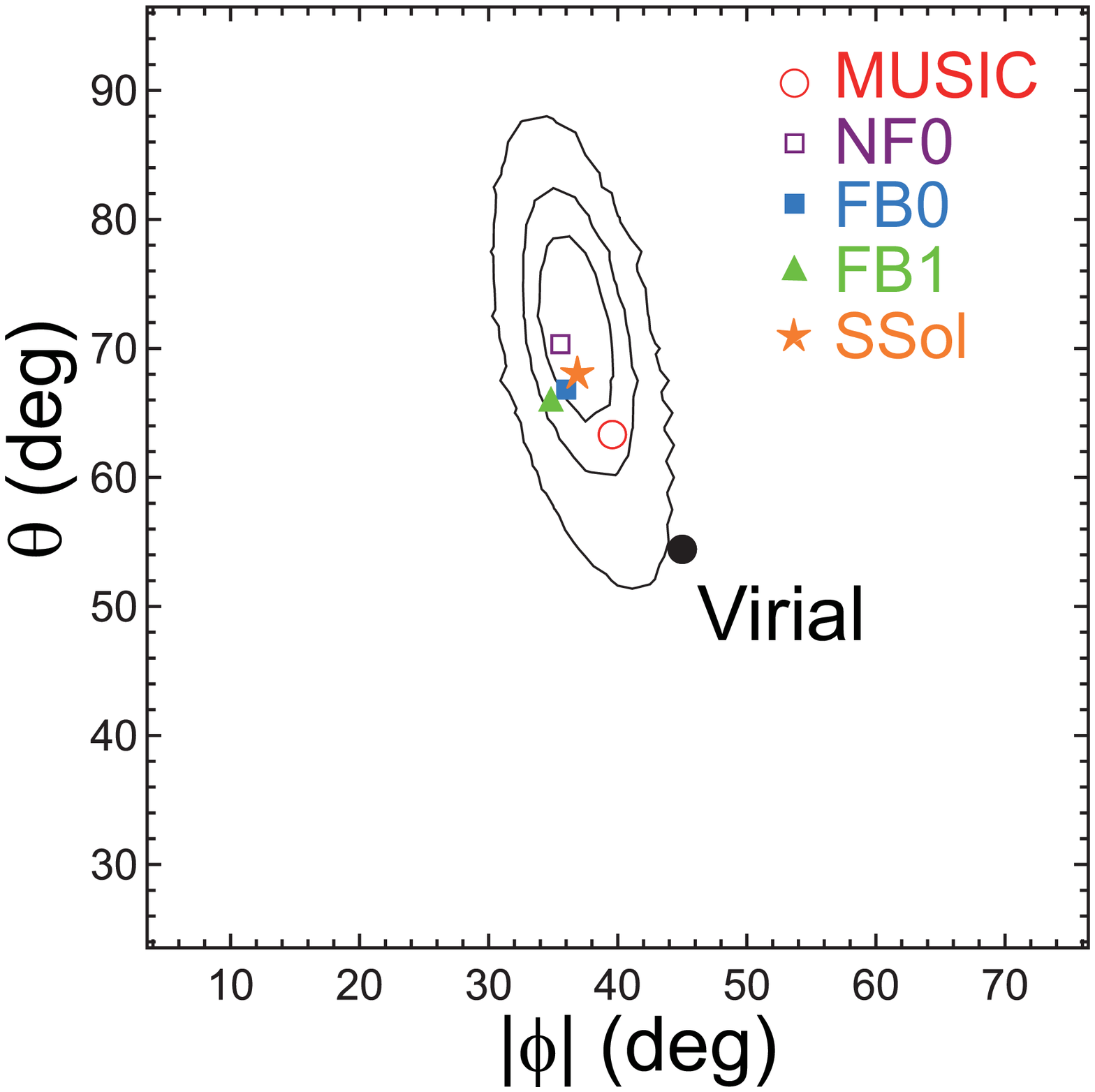} \caption{Direction of the plane normal
$P_3=(a,b,c)$ in the space of $(\log r_s, \log M_s, \log T_{\rm X})$;
$\theta$ is the angle between $P_3$ and the $\log T_{\rm X}$ axis, and
$\phi$ is the azimuthal angle around the $\log T_{\rm X}$ axis, measured
anti-clockwise from the $\log r_s$ axis, or $\tan\phi=b/a$.  Probability
contours are shown for the observed clusters at the 68 ($1\:\sigma$),
90, and 99\% confidence levels from inside to outside. The contours are
elongated in the direction of rotation around $P_1$
(Figure~\ref{fig:plane}), to which the direction $P_3$ is less
constrained. The prediction of the virial expectation ($r_s M_s^{-1}
T_{\rm X} \propto \rm const$) corresponds to $(\phi, \theta) =
(-45^\circ,55^\circ)$ (black dot), and is rejected at the $>99$\%
confidence level. Note that for the virial expectation the angle
$\theta$ is the one between vectors
$(1/\sqrt{3},-1/\sqrt{3},1/\sqrt{3})$ and $(0,0,1)$, which is $\approx
55^\circ$. The plane normals derived for simulation samples MUSIC, NF0,
FB0, and FB1 are shown by the open red circle, the open purple square,
the filled blue square, and the filled green triangle, respectively. All
the simulated angles are located inside the 90\% contour level and are
consistent with the observations at that level. A prediction based on a
similarity solution is shown by the orange star (SSol).\label{fig:prob}}
\end{figure}

The plane is described by $a\log r_s + b\log M_s + c\log T_{\rm
X}=\mathrm{const.}$, with $a=0.76^{+0.03}_{-0.05}$,
$b=-0.56^{+0.02}_{-0.02}$, and $c=0.32^{+0.10}_{-0.09}$. Likelihood
contours of the parameters describing the direction of the plane normal,
$P_3=(a,b,c)$, are shown in Figure~\ref{fig:prob}. The estimation of the
errors is described in the \ref{sec:append}ppendix. If the intracluster
gas at $r\lesssim r_s$ simply preserves its pressure equilibrium state
at the cluster formation, the gas temperature should reflect the
potential depth of the dark-matter halo at the formation.  Thus, one may
expect that the gas temperature follows the virial theorem in a narrow
sense (``virial expectation'') at that time, $T_{\rm X} \propto
M_s/r_s$, which is one of the main assumptions for the self-similar
scaling relations of clusters. The resulting plane, however, is
significantly tilted from this virial expectation (Figure~\ref{fig:prob})
and is represented by $T_{\rm X}\propto M_s^{-b/c} r_s^{-a/c}\propto
M_s^{1.8\pm 0.5}/r_s^{2.3\pm 0.7}$. Our findings show that the
temperature $T_{\rm X}$ is more sensitive to the depth of the
gravitational potential represented by $M_s/r_s$ than the canonical
virial expectation because $T_{\rm X}\propto
M_s^{-0.2}r_s^{-0.3}(M_s/r_s)^{2.0}$. In other words, clusters with a
deeper potential well tend to have higher temperatures $T_{\rm X}$ than
the virial expectation, or visa versa.

\begin{figure*}
\plottwo{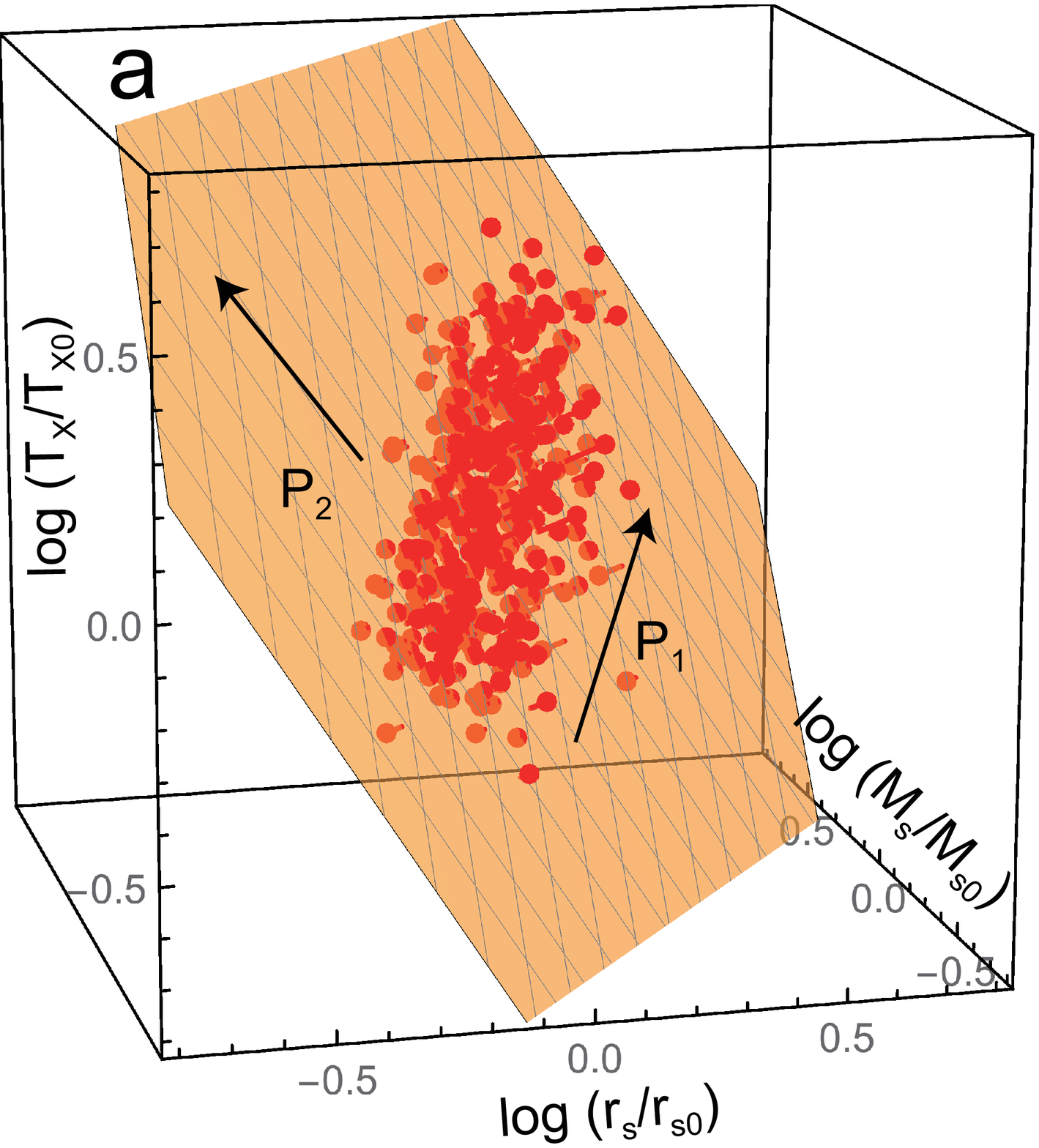}{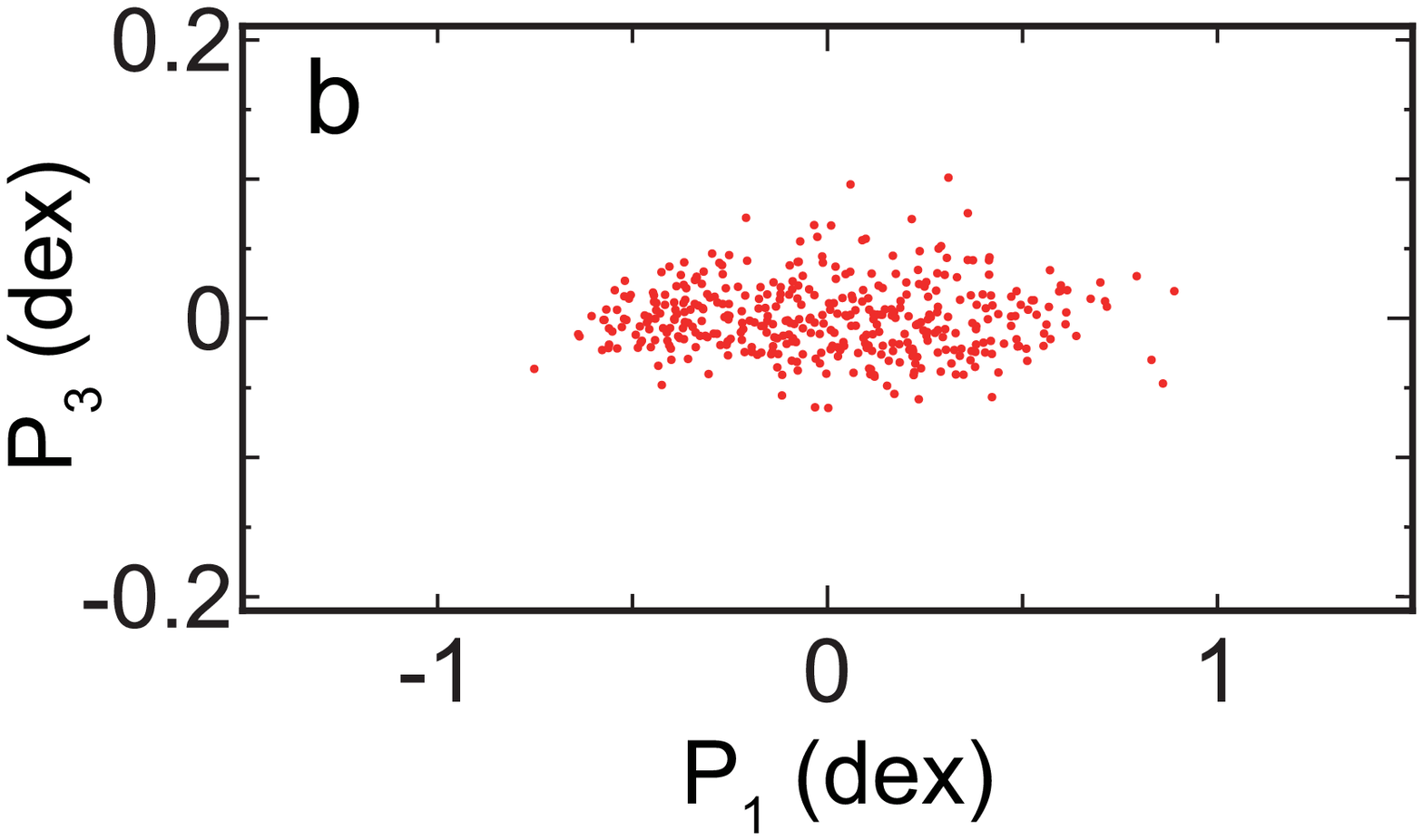} \caption{Same as
    Figures~\ref{fig:plane}(a) and (b) but for the MUSIC clusters. The
    axes are normalized by the average parameters of the sample
    ($r_{s0}=414$~kpc, $M_{s0}=1.4\times 10^{14}\: M_\odot$, and $T_{\rm
    X0}=3.7$~keV). The plane and the directions $P_1$, $P_2$, and $P_3$
    are determined for the MUSIC sample
    (Table~\ref{tab:vec}).\label{fig:planemusic}}
\end{figure*}

\section{Comparison with Numerical Simulations}

Our CLASH sample includes only 20 clusters, and our results may be
affected by observational biases. Here, we examine the results of
numerical simulations to properly interpret the observations in the
context of the CDM cosmology and discuss possible selection bias.
First, we analyzed the outputs of MUSIC $N$-body/hydrodynamical
simulations \citep{men14a}. These are adiabatic; that is, they do not
include any nongravitational effects such as feedback from active
galactic nuclei (AGNs) or from supernovae (SNe), and there is no
radiative cooling.

The details of the simulations for the MUSIC sample are given in
\citet{men14a}. The MUSIC sample is obtained by resimulating halos
selected from the MultiDark cosmological simulation
\citep{2012MNRAS.423.3018P} in order to achieve a higher resolution. The
parallel TREEPM+SPH GADGET code \citep{2005MNRAS.364.1105S} is used for
the resimulations. The mass resolution for the dark-matter particles is
$m_{\rm DM}=9.01\times 10^8\: h^{-1}\: M_\odot$ and that for the gas
particles is $m_{\rm SPH}=1.9\times 10^8\: h^{-1}\: M_\odot$, where the
Hubble constant is written as $H_0=100\: h$\,km\,s$^{-1}$\,Mpc$^{-1}$
and $h=0.7$. The gravitational softening is $6\: h^{-1}$~kpc for the
both gas and dark-matter particles in the high-resolution regions. We
select all of the 402 clusters at $z=0.25$ with $M_{200}>2\times
10^{14}\: h^{-1}\: M_\odot$ regardless of dynamical state, where
$M_{200}$ is the mass enclosed within a sphere of radius ($r_{200}$)
within which the mean overdensity equals 200 times the critical density
of the universe. We compute the mass-weighted temperature including the
core. The mass-weighted formulation is the most appropriate to evaluate
the thermal energy of the X-ray gas to be included in the virial
theorem. In addition, we kept the core because these simulations are
nonradiative and thus do not present cool-core features. The scale
radius $r_s$ is obtained by fitting the total density distribution
(gas+dark matter) with the NFW profile up to $r_{200}$. The mass $M_s$
is then derived as the mass enclosed by a sphere of radius $r_s$.

We see that 402 simulated MUSIC clusters at the redshift of $z=0.25$
form a plane in the $(\log r_s, \log M_s, \log T_{\rm X})$ space
(Figure~\ref{fig:planemusic}).  In Figure~\ref{fig:plane}(b), we project the
simulated clusters on the cross-section of the observed plane, showing
that the two sets of data are distributed around the same plane
($P_3=0$), although the band-like distribution of the MUSIC data is
slightly tilted (see Figure~\ref{fig:prob}) while passing through the
origin. Many of them are found at smaller $P_1$, because the average
radius and mass of the MUSIC clusters are smaller than those of the
observed clusters by a factor of a few. The plane angle is consistent
with the observed one (90\% confidence level) and deviates from the
virial expectation (Figure~\ref{fig:prob}). The dispersion around the
plane for the simulated clusters is $0.025$ dex and is even smaller than
the observed one ($0.045^{+0.008}_{-0.007}$ dex;
Table~\ref{tab:disp}). Since unrelaxed clusters tend to have disturbed
internal structure, they are expected to increase the
dispersion. However, even when we choose the 20\% most unrelaxed (UR)
clusters in the sample, it is only 0.033. Therefore, the slightly larger
observed thickness of the plane is unlikely ascribable to the dynamical
state of the systems. The direction of the plane for the UR clusters is
also not much different from that of the full sample
(Table~\ref{tab:vec}). Although our CLASH clusters are relatively
massive and relaxed, these results of the numerical simulations show
that the selection bias should not significantly affect the derived
plane parameters.

To study the evolution in detail, we analyzed another set of simulation
data \citep{ras15a}. Each of the samples named FB0 and NF0 consists of
29 massive clusters at $z=0$, and contain both relaxed and unrelaxed
ones. FB0 includes nongravitational effects and NF0 does not.  The
details of the simulations for samples FB0, FB1, and NB0 are given in
\citet{ras15a} and \citet{2017MNRAS.467.3827P}. They are also carried
out with the GADGET code \citep{2005MNRAS.364.1105S} but including an
updated SPH scheme \citep{2016MNRAS.455.2110B}. The simulations consist
in 29 Lagrangian regions around massive clusters with $M_{\rm 200}\sim
1$--$30\times 10^{14}\: h^{-1}\: M_\odot$ at $z=0$.  The simulations FB0
and FB1 include phenomena such as heating by AGNs and SNe in addition to
radiative cooling, while NF0 is from nonradiative runs. Samples FB0 and
NF0 consist of the clusters at $z=0$, while sample FB1 refers to the
runs at $z=1$. The mass resolution for the dark-matter particles is
$m_{\rm DM}=8.3\times 10^8\: h^{-1}\: M_\odot$ and that for the initial
gas particles is $m_{\rm SPH}=1.5\times 10^8\: h^{-1}\: M_\odot$. The
gravitational softening is $3.75\: h^{-1}$~kpc for both the gas and
dark-matter particles in the high-resolution regions
\citep{2017MNRAS.468..531B}. We derive $r_s$ and $M_s$ for the clusters
in the FB0, FB1, and NB0 samples using the same method exploited for the
MUSIC simulations. Since this sample, contrary to the previous sample,
is built on radiative simulations we do exclude the core. As we did for
the observed CLASH sample, the temperatures are obtained in the region
between 50 and 500 kpc from the cluster centers. Although we use the
mass-weighted temperature in the following discussion, we have confirmed
that the results such as the plane angle and thickness are not
significantly affected by the choice of the temperature weighting
(e.g. spectroscopic-like temperature; \citealt{maz04a}) or the choice of
the metric radius for temperature measurements ($> 500$~kpc).

\begin{figure*}
\plottwo{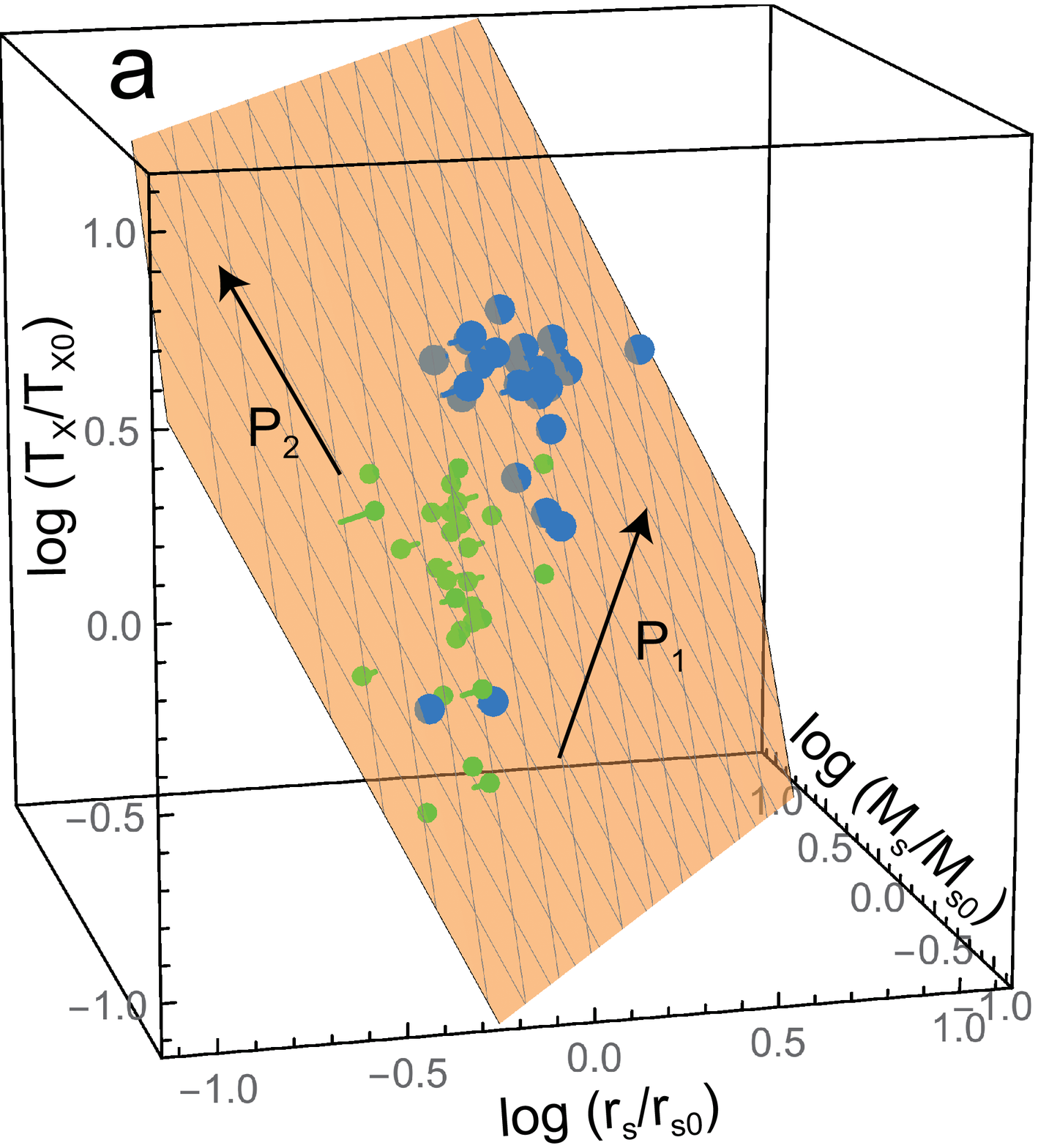}{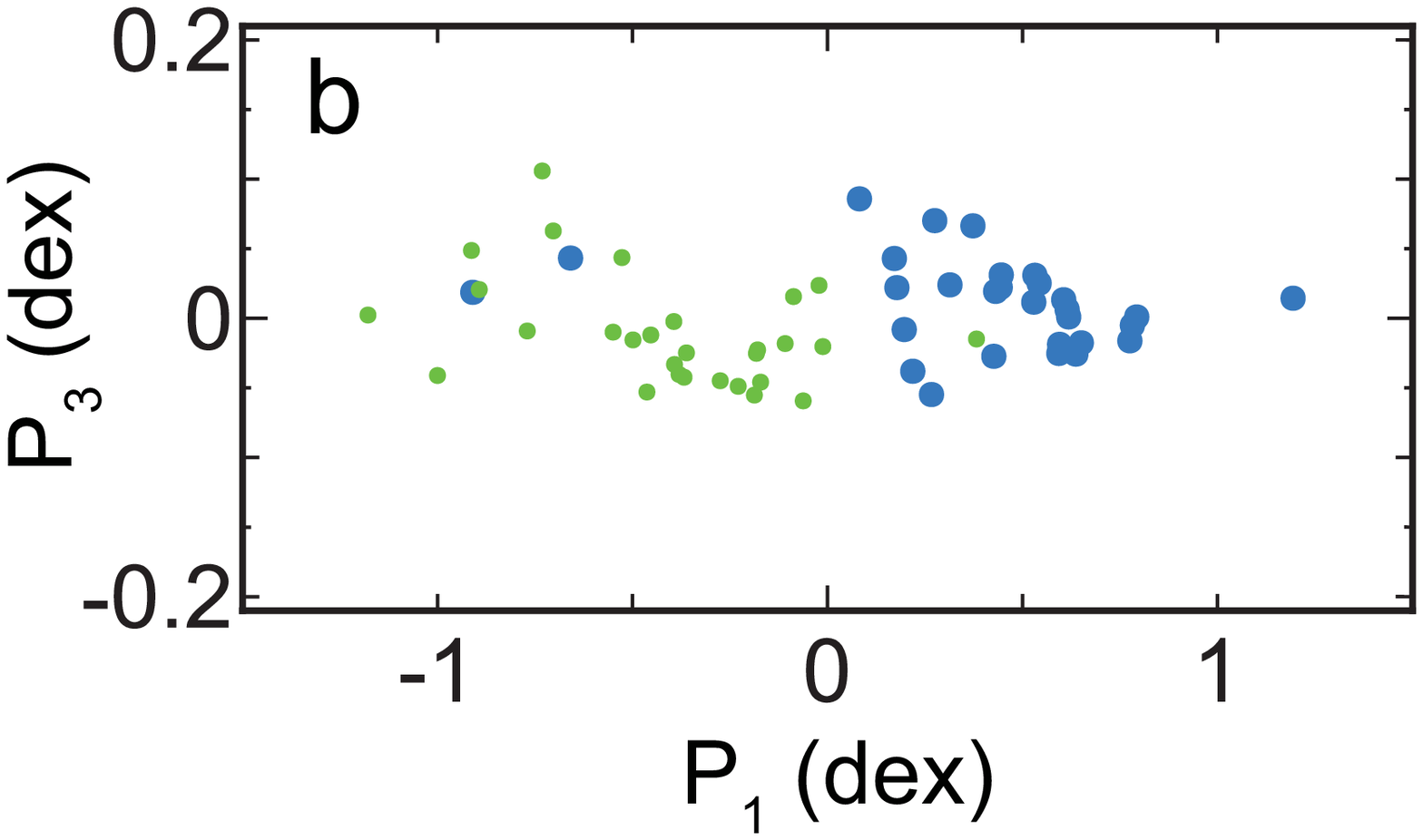} \caption{Same as
Figures~\ref{fig:plane}(a) and (b) but for the simulated clusters. Large
blue and small green points show the simulation samples FB0 ($z=0$) and
FB1 ($z=1$), respectively. The axes are normalized by the average
parameters of the combined sample (FB0+FB1; $r_{s0}=388$~kpc,
$M_{s0}=1.4\times 10^{14}\: M_\odot$, and $T_{\rm X0}=4.8$~keV). The
plane and the directions $P_1$, $P_2$, and $P_3$ are determined for the
combined sample (Table~\ref{tab:vec}).\label{fig:sim}}
\end{figure*}

We find that each sample forms a plane whose angle is consistent with
the observed one (Figure~\ref{fig:prob}). The plane angles for FB0 and NF0
are almost the same, which means that the result is independent of the
gas physics. The lack of dependence means that radiative cooling and SNe
and AGN feedback counterbalance one another with the effect of not
drastically changing the X-ray gas profile on a scale of $r_s$.  Thus,
even if our CLASH clusters are affected by some selection bias
originating from gas physics (e.g. difference of AGN activities), the
bias does not have a significant impact on the plane parameters. The
thickness of the plane is, however, increased by the nongravitational
effects. In fact, the dispersion around the plane for FB0 is $0.031$
dex, which is larger than that for NF0 ($0.023$ dex), but is still
smaller than the observed one even if the observational errors are
considered (Table~\ref{tab:disp}). In Figure~\ref{fig:prob}, we also show
the plane angle for clusters in the same simulation as FB0 but at $z=1$
(sample FB1). Most of the clusters (25/29) are the progenitors of those
in FB0. While more clusters should be in the fast-rate growth phase at
$z=1$, the plane angle is not much different from that at $z=0$
(FB0). We find that the clusters in the samples FB0 and FB1 are
virtually on the same plane (Figure~\ref{fig:sim}); though, FB1 clusters
tend to have smaller physical radii and masses. This indicates that the
redshifts of the clusters are unlikely to impact the plane
parameters. The dispersion around the common plane is $0.037$ dex
(FB0+FB1 in Table~\ref{tab:disp}). This indicates that the clusters
evolve on this unique plane along the direction of $P_1$, and that
the evolution of cluster halo structure and the thermodynamic history of
intracluster gas are strictly regulated by the plane.

In general, current numerical simulations are realistically reproducing
the observed scaling relations including their
slopes \citep{tru18a}. Any possible small discrepancy of the
normalizations does not significantly affect the plane angle and the
cluster evolution along the plane.

\begin{figure*}
\epsscale{.40}\plotone{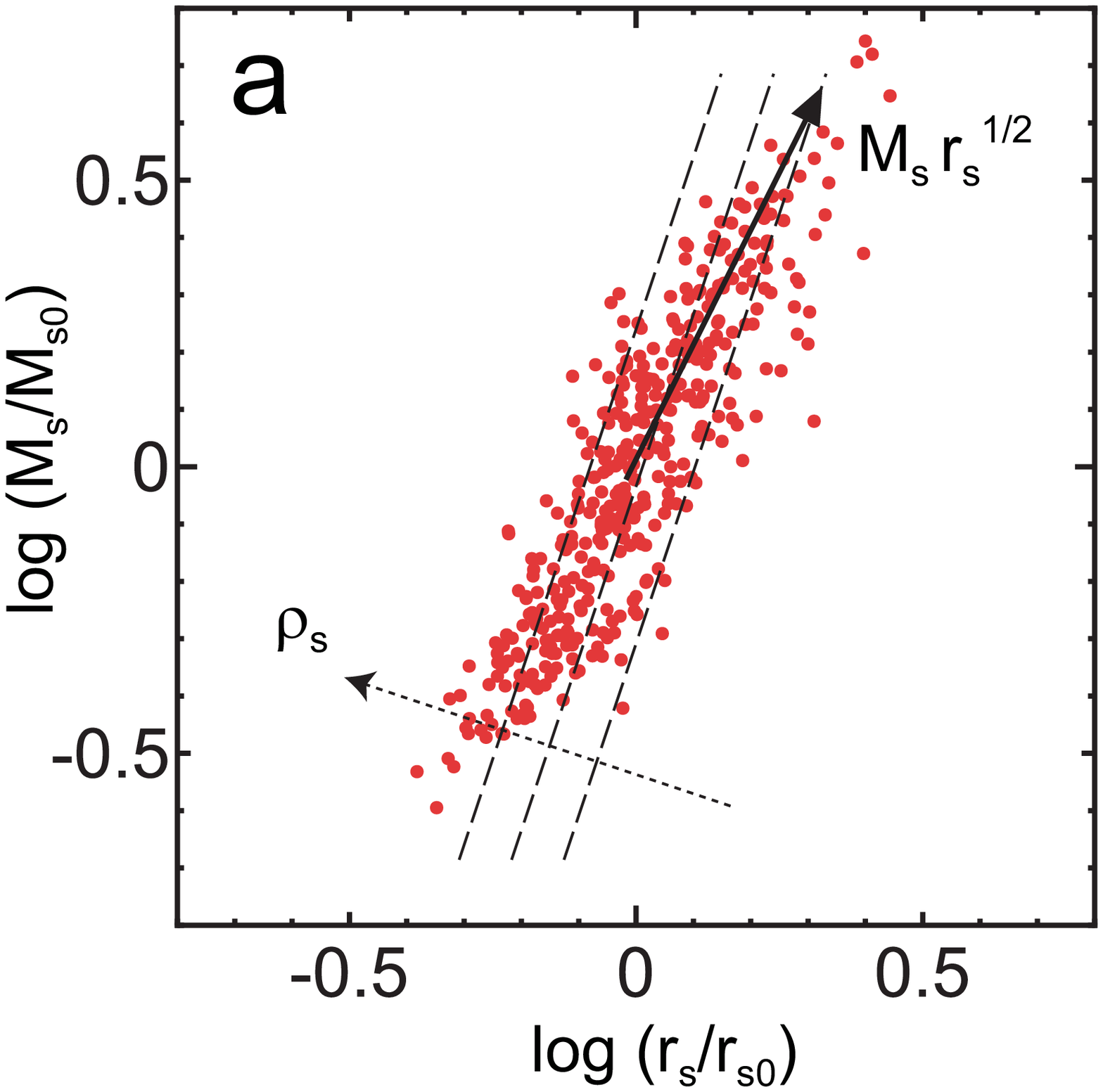}\\ \epsscale{.80}
\plottwo{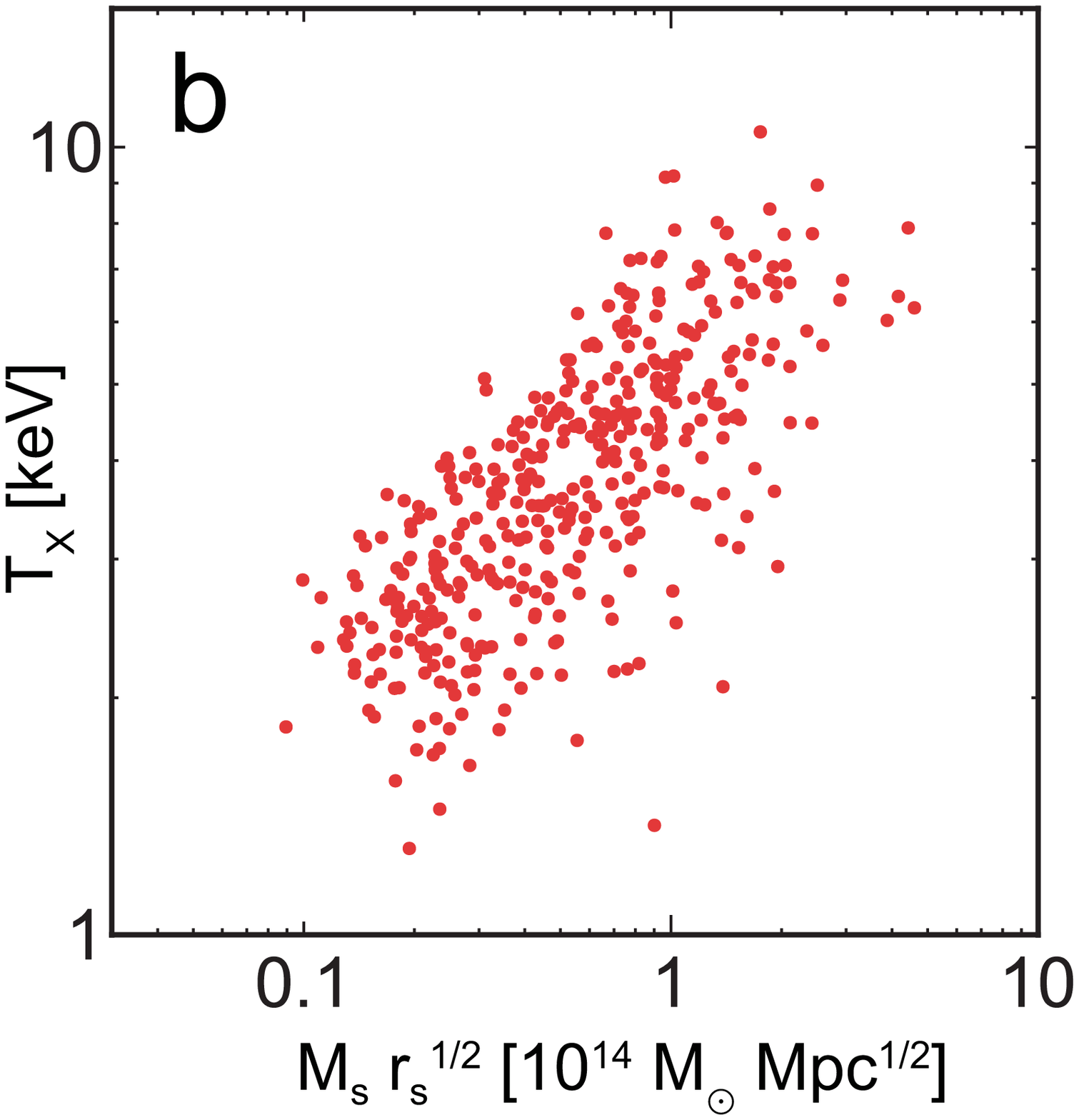}{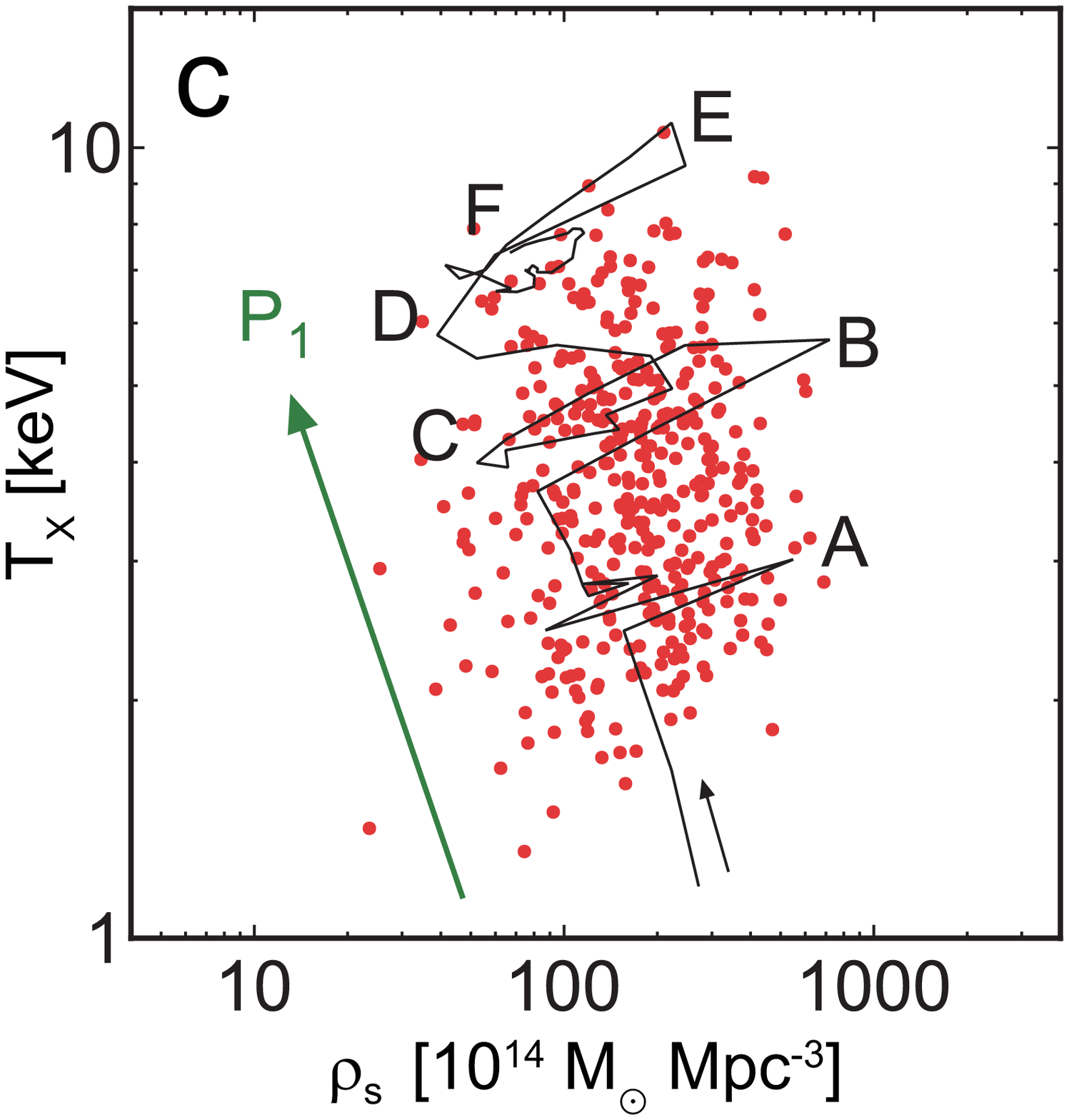} \caption{(a) Projection of the MUSIC data
points in Figure~\ref{fig:planemusic}(a) on the $\log r_s$--$\log M_s$
plane. The normalizations of the axes are the same as those in
Figure~\ref{fig:planemusic}. Each dashed line satisfies $\rho_s=\rm const$
and the value of the constant increases in the direction of the thin
dotted arrow. The thick solid arrow represents a line $r_s\propto
M_s^{1/2}$, and the value $M_s r_s^{1/2}$ increases in the
direction. (b) $T_{\rm X}$ is plotted against $M_s r_s^{1/2}$. (c) $T_{\rm X}$ is
plotted against $\rho_s$. Evolution of a typical cluster belonging to
FB0+FB1 is shown in the solid line. Prominent features are marked by the
labels A, B, ..., F. The cluster and the labels are the same as those in
Figure~\ref{fig:evol}. The green arrow shows the projected direction of
$P_1$ for the MUSIC data. \label{fig:musicrel}}
\end{figure*}

\begin{figure*}
\plottwo{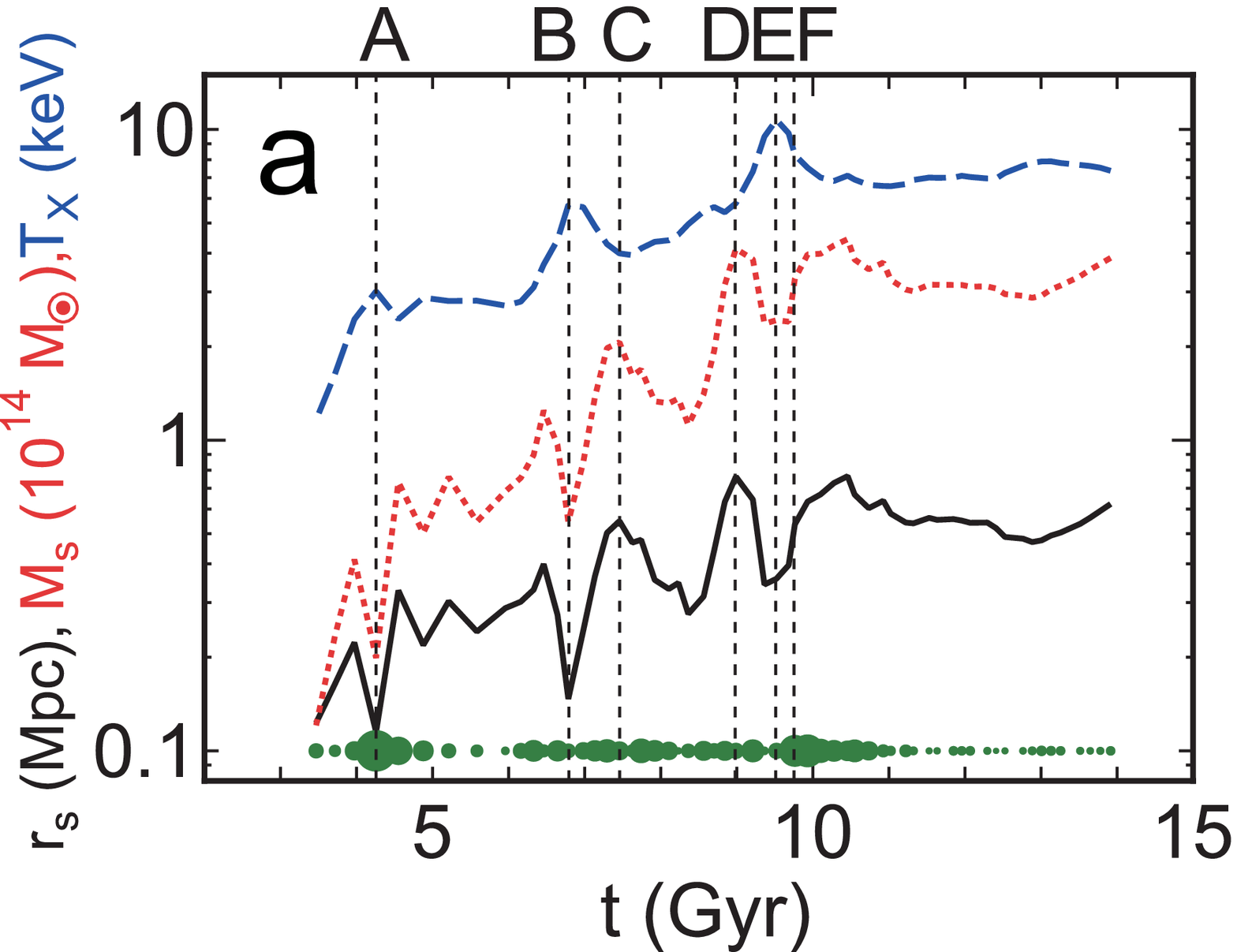}{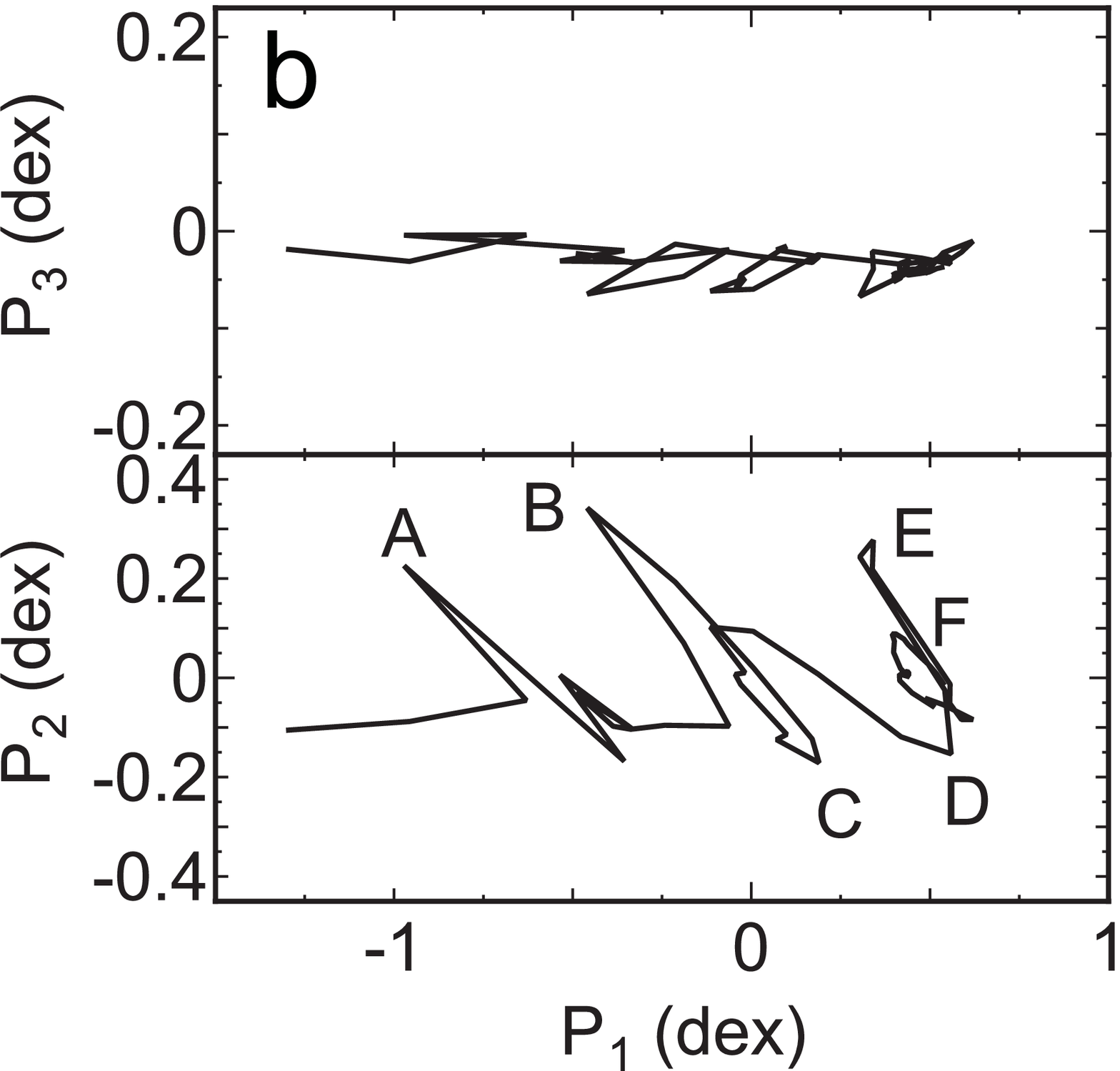}
\caption{(a) Evolution of $r_s$ (black solid), $M_s$ (red
    dotted), and $T_{\rm X}$ (blue dashed) of a typical cluster in the sample
    FB0+FB1. The time $t$ is the cosmological one. Prominent features
    are marked by the labels A, B, ..., F.  The area of the filled
    circles at the bottom is proportional to the reduced $\chi^2$ of the
    fit to the NFW profile. (b) Evolution of the cluster in the
    plane coordinate that is the same as that in Figure~\ref{fig:sim}. The
    labels A, B, ..., F correspond to those in (a). Note that scales
    of the axes $P_1$, $P_2$, and $P_3$ are
    different.\label{fig:evol}}
\end{figure*}

\section{Discussion}

\subsection{The plane angle predicted by a similarity solution and 
cluster evolution}
\label{sec:SSol}

In this subsection, we attempt to explain the origin of the peculiar
plane angle we found using an analytic solution.  \citet{ber85a}
constructed a one-dimensional similarity solution for secondary infall
and accretion onto an initially overdense perturbation in an Einstein-de
Sitter ($\Omega_0=1$) universe. For this solution, an object continues
to grow, and the matter density $\rho$, pressure $p$ at the radius $r$,
and mass $m$ inside $r$ at the cosmological time $t$ can be expressed by
nondimensional functions ($D$, $P$, and $M$):
\begin{eqnarray}
\label{eq:nond}
 \rho(r,t) &=& \rho_{\rm H} D(\lambda)\:,\nonumber
\\
p(r,t) &=& \rho_{\rm H}(r_{\rm ta}/t)^2 P(\lambda)\:,
\\
m(r,t) &=& (4\pi/3)\rho_{\rm H} r_{\rm ta}^3 M(\lambda)\:,\nonumber
\end{eqnarray}
where $r_{\rm ta}(t)$ is the maximum radius that a mass shell reaches
(the turnaround radius), $\rho_{\rm H}\propto t^{-2}$ is the density of
the background universe, and $\lambda=r/r_{\rm ta}$ is the
nondimensional radius. The turnaround radius is represented by $r_{\rm
ta}=A_{\rm ita}t^{8/9}$, where $A_{\rm ita}$ is the coefficient that
depends on the overdense perturbation \citep{ber85a}. The solution has
an entropy integral,
\begin{equation}
\label{eq:enti}
 P(\lambda) D(\lambda)^{-\gamma}
M(\lambda)^{10/3-3\gamma} = {\rm const}\:,
\end{equation}
where $\gamma=5/3$ is the adiabatic index. This relation holds even for
a system with a mixture of gas and dark matter \citep{ber85a}. From
equations~(\ref{eq:nond}) and~(\ref{eq:enti}), we have
$p\rho^{-5/3}m^{-5/3}\propto A_{\rm ita}^{-3}$, which does not depend on
$t$. The coefficient can be written as $A_{\rm ita}=r_{\rm ita}/t_{\rm
ita}^{8/9}$, where $r_{\rm ita}$ and $t_{\rm ita}$ are the turnaround
radius and time of the overdense perturbation, respectively. Note that
$t_{\rm ita}$ is much earlier than ``the formation of a cluster''
discussed in this study. Assuming that the evolution of the overdense
perturbation follows a theory of a spherical collapse, they are
represented by
\begin{equation}
\label{eq:ita}
 r_{\rm ita}\propto m_{\rm ita}^{(n+5)/6},\hspace{15mm} t_{\rm
ita}\propto m_{\rm ita}^{(n+3)/4}\:,
\end{equation}
where $m_{\rm ita}$ is the mass scale of the perturbation and $n$ is the
local slope of the primordial matter power spectrum\footnote{Note that
the relations are applied to the overdense perturbation and not to the
whole cluster.}  \citep{kai86a,pee93}, which is $n\sim -2$ at cluster
scales \citep{eis98a,die15a}. Thus, we obtain
$p\rho^{-5/3}m^{-5/3}\propto m_{\rm ita}^{-5/6}$. Assuming that
$p\propto \rho T_{\rm X}$ and $\rho\propto M_s/r_s^3$ at $r\sim r_s$,
the relation is $r_s^2 M_s^{-7/3}T_{\rm X}\propto m_{\rm ita}^{-5/6}$.
Here, we speculate that the structure of the NFW profile at $r\lesssim
r_s$ reflects the overdense perturbation that initially collapsed in the
similarity solution by \citet{ber85a}. In other words, the fast-rate
growth of a dark-matter halo is related to the initial collapse. In
fact, \citet{cor15b} demonstrated that the characteristic density
$\rho_s$ of the NFW profile is proportional to the critical density of
the background universe $\rho_c$ at the time when the dark-matter halo
transits from the fast-rate to the slow-rate growth phase (i.e. the halo
formation time). On the other hand, since the initial collapse can be
described as a simple spherical collapse of an overdense region, the
typical density of the collapsed object is also proportional to $\rho_c$
at the collapse time \citep{ber85a}. This indicates that both the inner
structure of the NFW profile and the overdense perturbation in the
similarity solution follow the same evolution and scaling
relation. Thus, we assume $M_s\propto m_{\rm ita}$ and $r_s \propto
r_{\rm ita}$, which leads to
\begin{equation}
\label{eq:enti2}
 r_s^2 M_s^{-3/2}T_{\rm X}={\rm const}\:.
\end{equation}
The angle of this plane is shown in Figure~\ref{fig:prob} (SSol) and it is
consistent with the observations. The essential point of the similarity
solution is that clusters are not isolated but continuously growing
through matter accretion from their outer environments. Therefore,
additional contributions, such as the flux of inertia at the cluster
surface, should be included in the virial theorem for the complete
description of the dynamical state \citep{ber85a}. In the $\Lambda$CDM
model \citep{pee93}, the cosmological constant becomes non-negligible at
$z\lesssim ((1-\Omega_m)/\Omega_m)^{1/3}-1\sim 0.39$, which is close to
the median redshift of our observational sample
(Table~\ref{tab:disp}). Thus, about half of the sample may be affected
by the cosmological constant. However, although the density profiles of
these objects may become steeper in the outskirts \citep{ber85a}, the
effect is not serious because we are interested in the inner region
($r\lesssim r_s$).

This simple model may also explain the vector $P_1$. Since we assumed
$r_s \propto r_{\rm ita}$ and $M_s \propto m_{\rm ita}$, we obtain $r_s
\propto M_s^{1/2}$ from equation~(\ref{eq:ita}). The direction of the
line $r_s \propto M_s^{1/2}$ on the $\log r_s$--$\log M_s$ plane is
almost the same as that of $P_1$ projected on the $\log r_s$--$\log M_s$
plane, especially for $P_1$ for the simulation samples
(Table~\ref{tab:vec}). Figure~\ref{fig:musicrel}(a) shows that the MUSIC
data points are actually distributed along the line $r_s \propto
M_s^{1/2}$ (thick solid arrow) on the $\log r_s$--$\log M_s$ plane. This
means that the cluster's elongated distributions along $P_1$
(Figures~\ref{fig:planemusic} and~\ref{fig:sim}) reflect the evolution
of the typical overdense perturbation along $r_{\rm ita} \propto m_{\rm
ita}^{1/2}$, which can also be interpreted as the evolution of clusters
during the fast-rate growth phase. The direction of $P_1$ for the
observational sample (first line of Table~\ref{tab:vec}) is slightly
different from those for the simulations. This may be because the
observational sample is biased toward high temperature clusters.

Cluster formation time is associated with the characteristic density
\citep[e.g.][]{fuj99d,lud13a,cor15b}. The dashed lines in
Figure~\ref{fig:musicrel}(a) are isochrones or $\rho_s = \rm const$. The
MUSIC data points are widely distributed along each isochrone, which
reflects a variety of cluster masses for a given formation time. In
other words, it reflects a variety of peak densities of initial density
fluctuations of the universe. Individual clusters evolve approximately
in the direction of $P_1$ (Figures~\ref{fig:planemusic}
and~\ref{fig:sim}), or in the direction to which $M_s r_s^{1/2}$
increases (Figure~\ref{fig:musicrel}(a)), that is, along the line $r_s
\propto M_s^{1/2}$ ($r_s \propto M_s^{1/1.65}$ in a detailed study;
\citealt{zha09a}).  Figure~\ref{fig:musicrel}(b) shows that $M_s
r_s^{1/2}$ and $T_{\rm X}$ are correlated, which reflects that $T_{\rm
X}$ evolves according to the structure evolution of dark halos, although
the projection of the band-like distribution of clusters in
Figure~\ref{fig:planemusic}(a) onto the $M_s r_s^{1/2}$--$T_{\rm X}$
plane disperses the relation. In Figure~\ref{fig:musicrel}(a), clusters
become denser (having larger $\rho_s$) or older in the direction of the
thin dotted arrow. In Figure~\ref{fig:musicrel}(c), the correlation
between $\rho_s$ and $T_{\rm X}$ is not clear, because a possible
correlation is obscured by the projection of clusters with various
masses along the dashed lines in Figure~\ref{fig:musicrel}(a) (see also
Figure~\ref{fig:planemusic}(a)). However, each cluster moves mainly
along the direction of $P_1$, which shows that the cluster temperature
increases as the density $\rho_s$ decreases, although major mergers (A,
B, and E in Figure~\ref{fig:musicrel}(c)) derail the cluster
significantly from the evolution along $P_1$. We emphasize that the
MUSIC data distribution on the plane does not follow a single line but
has a finite spread, and $M_s$ has a distribution for a given formation
time.  In general, the correlation between some two parameters of
clusters is not necessarily represented by a line but is often
represented by a broad band (Figure~\ref{fig:musicrel}). This is because
the mass scale of the initial density fluctuation of the universe has a
distribution for a given peak height \citep[e.g.][]{bar01a}. Since the
peak height also has its own distribution, these result in a
two-dimensional band-like distribution of clusters.  In general,
clusters move along the band on the plane in the direction of $P_1$ (see
also the next subsection).

\begin{figure*}
\plottwo{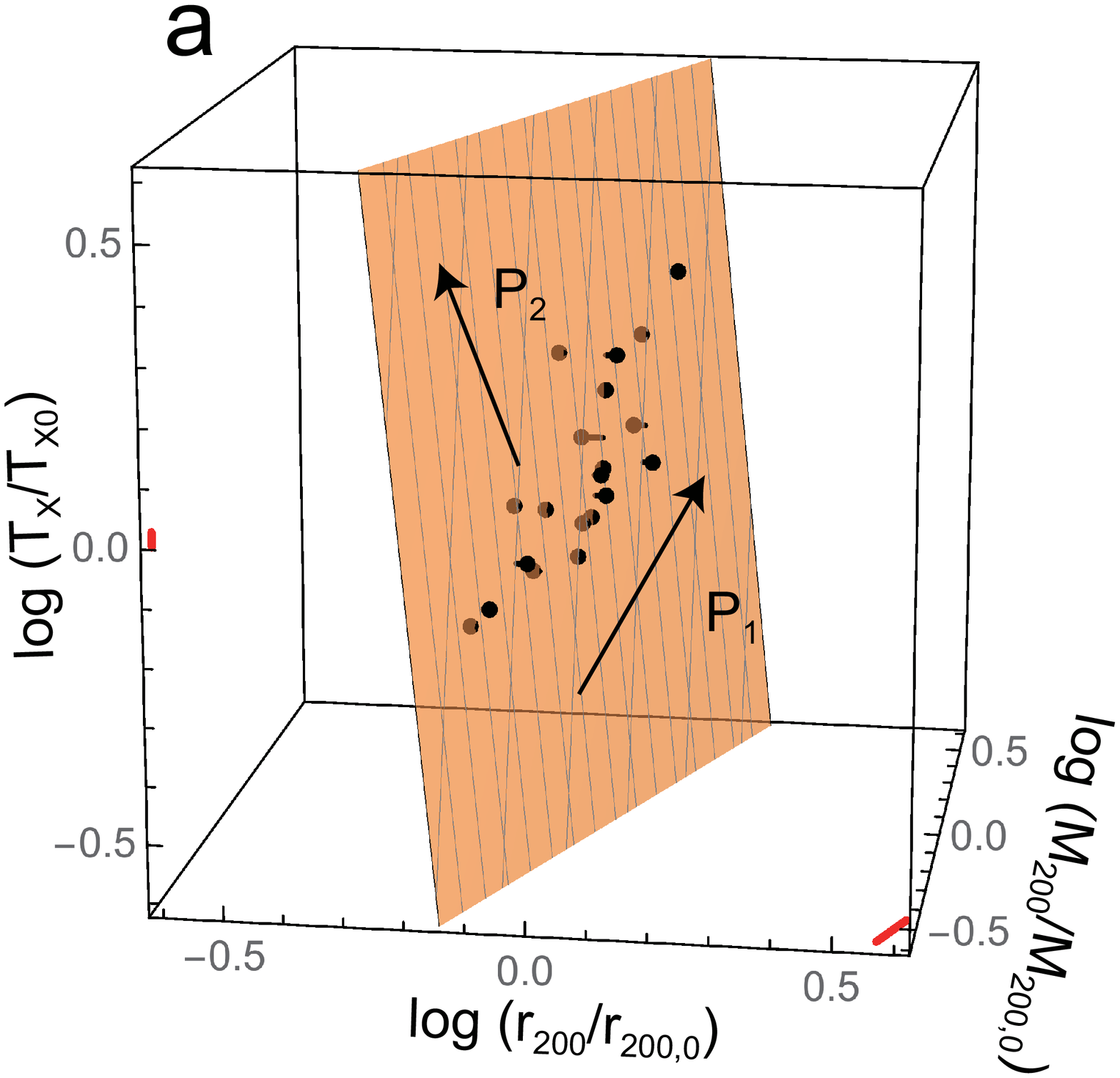}{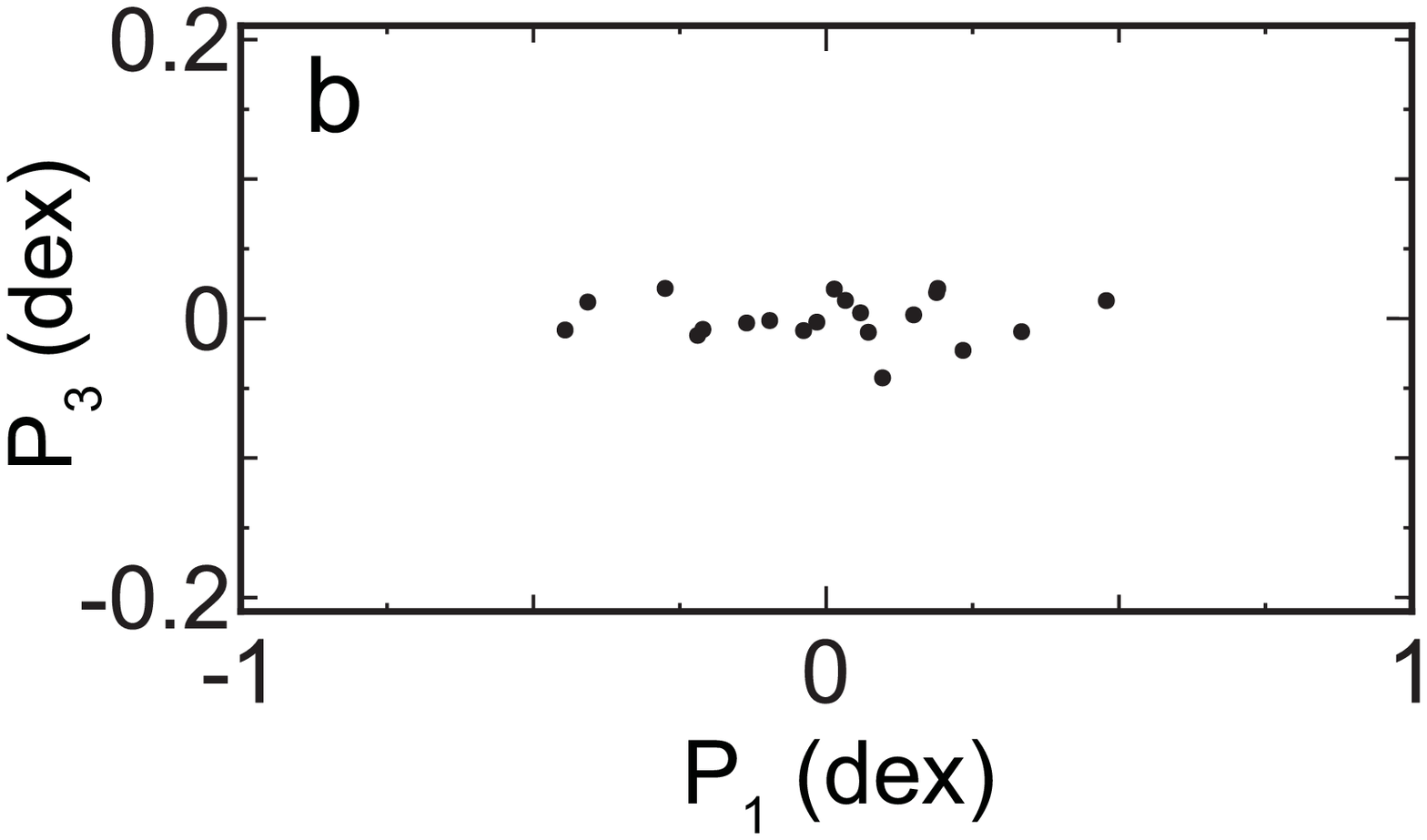} \caption{Same as
Figures~\ref{fig:plane}(a) and (b), but in the space of $(\log
(r_{200}/r_{200,0}), \log (M_{200}/M_{200,0}), \log (T_{\rm X}/T_{\rm
X0}))$, where $r_{200,0}=2040$~kpc, $M_{200,0}=1.4\times 10^{15}\:
M_\odot$, and $T_{\rm X0}=8.2$~keV are the sample averages of $r_{200}$,
$M_{200}$, and $T_{\rm X}$, respectively.\label{fig:plane200}}
\end{figure*}

\subsection{Stability of the plane against mergers}

The similarity of clusters is generally good because most of them are
well represented by the NFW profile. The thinness of the plane reflects
the excellent similarity because cluster structure is well described by
the similarity solution discussed in 
Section~\ref{sec:SSol}. However, clusters occasionally experience
mergers that might break the similarity.

In Figure~\ref{fig:evol}(a), we follow the evolution of a cluster that
undergoes three major mergers in the sample FB0+FB1. This cluster
experiences mergers around the times indicated by A, B, and E. While
$M_s$ tends to correlate with $r_s$, the temperature $T_{\rm X}$ tends
to inversely correlate with $r_s$. This can be explained as
follows. When a large substructure is merging with a cluster, it does
not dissolve immediately when it touches the viral radius of the
cluster. The 3D cluster gas density profiles will, therefore, include
the gas of the substructure as it moves within the cluster external
atmosphere toward the inner regions. Very large substructures can even
reach the center and cross it. This is the main origin of changing $r_s$
and $M_s$: when the substructure is in the outskirts, the profile is
flatter and thus $r_s$ and $M_s$ are larger (less concentrated cluster),
while when the substructure is closer to the center, the object appears
more concentrated, i.e. with smaller $r_s$ and $M_s$. On the other hand,
$T_{\rm X}$ increases for a moment, after the shock has time to
propagate throughout a large region of the cluster.

Thanks to the behavior of the three parameters $r_s$, $M_s$, and $T_{\rm
X}$, the cluster does not substantially deviate from the plane ($P_3=0$)
even during a merger (Figure~\ref{fig:evol}(b)), which contributes to
the thinness of the plane.  Figure~\ref{fig:evol} also shows that after
the end of the last major merger indicated by the letter F, the three
parameters do not change much and the cluster remains in almost the same
position on the plane. This indicates that the current cluster structure
is determined at the last major merger.

\subsection{Cluster distribution in the space of $(\log r_{200}, \log
M_{200}, \log T_{\rm X})$.}  \label{sec:200}

Contrary to the inside-out scenario, dark-matter halos possess only a
common global density that is related to the virial overdensity in the
classical picture of cluster formation
\citep{gun72a,1974ApJ...187..425P,lac93a}. A representative density is
$\rho_{200}$, which is 200 times the critical density of the universe at
the cluster redshift. Individual halos can thus be characterized by
$M_{200}$, the mass enclosed within a sphere of radius $r_{200}$ within
which the mean overdensity equals $\rho_{200}$. The mass $M_{200}$ is
often regarded as the total mass of a cluster. The values of $r_{200}$
and $M_{200}$ for our observed clusters are shown in
Table~\ref{tab:data}.

As an alternative parameter combination, we check the cluster
distribution in the space of $(\log r_{200}, \log M_{200}$, $ \log
T_{\rm X})$.  In Figure~\ref{fig:plane200}(a), the data points are
distributed on a plane almost vertical to the $\log r_{200}$--$\log
M_{200}$ plane, which reflects the obvious relation of $M_{200}=4\pi
\rho_{200} r_{200}^3/3 \propto r_{200}^3$ regardless of $T_{\rm X}$. The
redshift dependence of $\rho_{200}$ is small because the majority of the
clusters in our sample are distributed in a relatively narrow range at
low-intermediate redshifts, although the small redshift dependence
slightly slants the the plane. The dispersion in the direction of the
plane normal ($P_3$) is small (Figure~\ref{fig:plane200}(b);
$0.016^{+0.001}_{-0.001}$ dex). A weak correlation ($P_1$), in which
clusters with a larger $M_{200}$ tend to have a larger $T_{\rm X}$, is
seen on the plane, but the dispersion ($P_2$) is relatively large
(Figure~\ref{fig:plane200}(a); $0.099^{+0.011}_{-0.011}$ dex). Although
this correlation $P_1$ has been studied as a mass--temperature relation
\citep[e.g.][]{sun09a,2014ApJ...794...48C}, the large dispersion
suggests that it is not a primary relation, and that the combination of
parameters ($r_s, M_s, T_{\rm X}$) rather than $(r_{200}, M_{200},
T_{\rm X})$ is appropriate for studying the connection between the X-ray
gas and the dark halo structure and the interplay between gas heating
and gravitational collapse in detail. The well-known large scatter of
the concentration ($c_{200}\equiv r_{200}/r_s$)--mass ($M_{200}$)
relation \citep[e.g.][]{2001MNRAS.321..559B} may be because $r_s$ has
less to do with $r_{200}$ and $M_{200}$.

\section{Conclusions}

In this study, we showed that observational data of high-mass galaxy
clusters form a plane in the three-dimensional logarithmic space of
their characteristic radius $r_s$, mass $M_s$, and X-ray temperature
$T_{\rm X}$ with a very small orthogonal scatter.  Since the evolution
history of a cluster is encoded in $r_s$ and $M_s$, the tight
correlation suggests that the gas temperature was determined at a
specific cluster formation time. We also found that the plane is tilted
with respect to $T_{\rm X} \propto M_s/r_s$, which is the plane expected
in the case of simplified virial equilibrium. This strange plane angle
can be explained by a similarity solution, which indicates that clusters
are not isolated but continuously growing through matter accretion from
their outer environments. In other words, the effects of the growth must
be considered when the internal structure of clusters is discussed. We
have shown that numerical simulations reproduce the observed plane and
its angle, regardless of the gas physics implemented in the code. The
simulations show that clusters evolve along the plane and they do not
deviate much from the plane even during major mergers, which contributes
to the overall thinness of the plane.

Further work is needed to understand the relation between the similarity
solution and the formation of the NFW halo structure
\citep{1998ApJ...499..542S,2004ApJ...604...18W}. Large-scale
high-resolution numerical simulations from high to low redshifts would
be useful to study these topics. It would also be interesting to extend
the observational sample to objects other than massive clusters in the
nearby universe. Those objects include clusters at higher redshifts and
less massive systems, such as galaxy groups and elliptical galaxies that
present a break from the cluster self-similarity behavior
\citep{pon99a}. It would also be interesting to study
relationships between the cluster plane and galactic-scale ones
(e.g. \citealt{gou16a}). Galaxy clusters have been used to constrain
cosmological parameters, such as the amount of matter and dark energy,
and to investigate the growth of large-scale structure. In particular,
the evolution of the abundance of rare massive clusters above a given
mass threshold is highly sensitive to the expansion history and the
growth rate of mass density fluctuations
\citep[e.g.][]{2009ApJ...692.1060V}. Thanks to the thinness of the plane
we discovered, precise determinations of the fundamental plane can be
used to calibrate cluster mass--observable relations, a key ingredient
of the cluster cosmology \citep{fuj18b}.

\acknowledgments

This work was supported by MEXT KAKENHI No.~15K05080
(YF). K.U. acknowledges support from the Ministry of Science and
Technology of Taiwan through the grants MoST 103-2112-M-001-030-MY3 and
MoST 106-2628-M-001-003-MY3. E.R. acknowledge support from the ExaNeSt
and EuroExa projects, funded by the European Union’s Horizon 2020
research and innovation programme under grant agreements No 671553 and
No 754337, respectively.

\clearpage

\begin{deluxetable*}{ccccccc}[h]
\tablecaption{Cluster data\label{tab:data}}
\tablenum{1}
\tablewidth{0pt}
\tablehead{
\colhead{Cluster} & \colhead{$z$} &
\colhead{$r_s$} &
\colhead{$r_{200}$} & \colhead{$M_s$} & \colhead{$M_{200}$} & 
\colhead{$T_{\rm X}$}\\
\colhead{} & \colhead{} & \colhead{(kpc)} &
\colhead{(kpc)} & \colhead{($10^{14}\: M_\odot$) } & 
\colhead{($10^{14}\: M_\odot$)} & \colhead{(keV)}
}
\startdata
\input{table1}
\enddata
\end{deluxetable*}

\begin{deluxetable*}{ccccccc}[ht]
\tablecaption{Cluster Samples\label{tab:disp}}
\tablenum{2}
\tablewidth{0pt}
\tablehead{\colhead{} &
\colhead{Observation} &
 \multicolumn{5}{c}{Simulations}\\
  \cline{3-7}
\colhead{} & \colhead{} & \colhead{MUSIC} & 
\colhead{NF0} &
\colhead{FB0} & \colhead{FB1} &
\colhead{FB0+FB1}
}
\startdata
Nongravitational effects & \nodata & no & no & yes & yes & yes \\
Redshift   & $0.377^{+0.309}_{-0.190}$ & 0.25 & 0 & 0 & 1 & $0+1$ \\
    Dispersion Around the Plane (dex) & $0.045^{+0.008}_{-0.007}$ 
& 0.025\tablenotemark{a} & 0.023 & 0.031 & 0.035 & 0.037\\
\enddata
\tablenotetext{a}{For the 20\% most relaxed (RE) and unrelaxed (UR)
clusters in the sample, it is 0.015 and 0.033, respectively. The
classification is based on fit residuals to the NFW profile.}
\end{deluxetable*}

\begin{deluxetable*}{cccc}[ht]
\tablecaption{Plane vectors\label{tab:vec}}
\tablenum{3}
\tablewidth{0pt}
\tablehead{
\colhead{Sample} &
\colhead{$P_1$} & \colhead{$P_2$} & \colhead{$P_3$}}
\startdata
Observation & 
$(0.55^{+0.03}_{-0.02},0.82^{+0.01}_{-0.01},0.15^{+0.04}_{-0.06})$ &
$(-0.34^{+0.08}_{-0.07},0.07^{+0.07}_{-0.07},0.93^{+0.03}_{-0.03})$ &
$(0.76^{+0.03}_{-0.05},-0.56^{+0.02}_{-0.02},0.32^{+0.10}_{-0.09})$ \\
MUSIC &
$(0.40,0.81,0.42)$ &
$(-0.60,-0.11,0.79)$ &
$(0.69,-0.57,0.44)$ \\
MUSIC (RE)\tablenotemark{a}&
$(0.38,0.82,0.42)$ &
$(-0.62,-0.11,0.78)$ &
$(0.69,-0.56,0.47)$ \\
MUSIC (UR)\tablenotemark{a}&
$(0.39,0.79,0.48)$ &
$(-0.58,-0.19,0.79)$ &
$(0.71,-0.59,0.39)$ \\
NF0 &
$(0.43,0.83,0.36)$ &
$(-0.48,-0.13,0.87)$ &
$(0.77,-0.55,0.34)$ \\
FB0 &
$(0.43,0.84,0.34)$ &
$(-0.51,-0.08,0.86)$ &
$(0.74,-0.54,0.39)$ \\
FB1 &
$(0.37,0.84,0.40)$ &
$(-0.55,-0.16,0.82)$ &
$(0.75,-0.52,0.40)$ \\
FB0+FB1 &
$(0.42,0.82,0.40)$ &
$(-0.53,-0.13,0.83)$ &
$(0.74,-0.56,0.38)$ \\
Observation\tablenotemark{b} &
$(0.27^{+0.01}_{-0.01},0.90^{+0.02}_{-0.03},0.35^{+0.06}_{-0.07})$ &
$(-0.18^{+0.03}_{-0.02},-0.31^{+0.06}_{-0.07},0.93^{+0.02}_{-0.02})$ &
$(0.94^{+0.01}_{-0.00},-0.32^{+0.01}_{-0.01},0.08^{+0.02}_{-0.02})$\\
\enddata
\tablenotetext{a}{The values for the 20\% most relaxed (RE) and unrelaxed (UR)
clusters in the MUSIC sample. The classification is based on fit
residuals to the NFW profile.}
\tablenotetext{b}{For the plane obtained in
the space of $(\log r_{200}, \log M_{200}, \log T_{\rm X})$ (see 
section~\ref{sec:200})}
\end{deluxetable*}

\clearpage

\appendix

\section{PCA and Error estimation}
\label{sec:append}

The plane that represents the planarly distributed data points in
three-dimensional space can be obtained so that the deviations of the
points from the plane are minimized. Here, we define three vectors
(principal components) in the space of $(x,y,z)\equiv (\log r_s, \log
M_s, \log T_{\rm X})$. The first component $P_1$ is defined as the
direction to which the points have the largest variance. The second
component $P_2$ is orthogonal to $P_1$ and is the direction to which the
points have the largest variance under the orthogonal condition between
$P_1$ and $P_2$. The third component $P_3$ is orthogonal both to $P_1$
and $P_2$, which means that the points have the least variance to that
direction. That is, $P_3$ is the normal to the plane that represents the
planarly distributed points.

We find the plane through a PCA. Assuming that each data point is given
by $(x_i, y_i, z_i)$, a covariant matrix can be defined as
\[
 A = \sum_i \left(
    \begin{array}{ccc}
      (x_i - \bar{x})^2 & (x_i - \bar{x})(y_i - \bar{y}) 
    & (x_i - \bar{x})(z_i - \bar{z}) \\
      (x_i - \bar{x})(y_i - \bar{y}) &  (y_i - \bar{y})^2 
    & (y_i - \bar{y})(z_i - \bar{z}) \\
      (x_i - \bar{x})(z_i - \bar{z}) & (y_i - \bar{y})(z_i - \bar{z})
    & (z_i - \bar{z})^2
    \end{array}
  \right)\:,
\]
where $\bar{x}$, $\bar{y}$, and $\bar{z}$ are the average of $x_i$,
$y_i$, and $z_i$, respectively. If the eigen values for the matrix $A$
are designated as $\lambda_1$, $\lambda_2$, and $\lambda_3$
($\lambda_1>\lambda_2>\lambda_3$), the corresponding eigen vectors are
the principal components $P_1$, $P_2$, and $P_3$,
respectively. The dispersions of data in the directions of $P_1$,
$P_2$, and $P_3$ are represented by $\sqrt{\lambda_1}$,
$\sqrt{\lambda_2}$, and $\sqrt{\lambda_3}$, respectively.

For the observational sample, we estimate the uncertainties of the plane
parameters ($P_1$, $P_2$, $P_3$, $\lambda_1$, $\lambda_2$, $\lambda_3$)
by accurately propagating the errors in the observed three cluster
parameters ($r_s$, $M_s$, $T_{\rm X}$) using Monte-Carlo simulations.
Among the three parameters ($r_s$, $M_s$, $T_{\rm X}$), the errors in
the NFW halo parameters ($r_s, M_s$) are highly correlated with each
other. Thus, when we estimate the errors, we directly use the joint
posterior probability distribution $P(M_{200}, c_{200})$ of the NFW
parameters for each cluster, accounting for the full uncertainty in both
mass and concentration. The posterior probability distributions for the
individual clusters were obtained by \citet{ume16a} using Markov-chain
Monte-Carlo sampling. We perform Monte-Carlo realizations of the
temperature $T_{\rm X}$ assuming random Gaussian errors. Finally, we
generate a total of $10^6$ realizations of the data and derive errors of
the plane parameters.

\end{document}

%% file: table1.tex
Abell~383         &
0.187 &
$   304
^{+  159}_{  -97}$&
$  1800
^{+  209}_{ -189}$&
$  1.4
^{+  1.0}_{ -0.5}$&
$  7.9
^{+  3.1}_{ -2.2}$&
$  6.5
\pm 0.24$
\\
Abell~209         &
0.206 &
$   834
^{+  243}_{ -192}$&
$  2238
^{+  161}_{ -172}$&
$  5.2
^{+  2.2}_{ -1.6}$&
$ 15.4
^{+  3.6}_{ -3.3}$&
$  7.3
\pm 0.54$
\\
Abell~2261        &
0.224 &
$   682
^{+  232}_{ -170}$&
$  2542
^{+  192}_{ -188}$&
$  5.8
^{+  2.7}_{ -1.8}$&
$ 22.9
^{+  5.6}_{ -4.7}$&
$  7.6
\pm 0.30$
\\
RX~J2129.7+0005   &
0.234 &
$   294
^{+  133}_{  -89}$&
$  1626
^{+  163}_{ -154}$&
$  1.1
^{+  0.7}_{ -0.4}$&
$  6.1
^{+  2.0}_{ -1.6}$&
$  5.8
\pm 0.40$
\\
Abell~611         &
0.288 &
$   560
^{+  250}_{ -172}$&
$  2189
^{+  204}_{ -208}$&
$  3.8
^{+  2.3}_{ -1.5}$&
$ 15.6
^{+  4.8}_{ -4.0}$&
$  7.9
\pm 0.35$
\\
MS~2137-2353      &
0.313 &
$   784
^{+  557}_{ -357}$&
$  2064
^{+  261}_{ -286}$&
$  4.7
^{+  5.2}_{ -2.6}$&
$ 13.4
^{+  5.8}_{ -4.9}$&
$  5.9
\pm 0.30$
\\
RX~J2248.7-4431   &
0.348 &
$   643
^{+  422}_{ -246}$&
$  2267
^{+  282}_{ -261}$&
$  4.9
^{+  4.8}_{ -2.3}$&
$ 18.5
^{+  7.8}_{ -5.7}$&
$ 12.4
\pm 0.60$
\\
MACS~J1115.9+0129 &
0.352 &
$   738
^{+  249}_{ -196}$&
$  2186
^{+  161}_{ -174}$&
$  5.1
^{+  2.4}_{ -1.7}$&
$ 16.6
^{+  4.0}_{ -3.7}$&
$  8.0
\pm 0.40$
\\
MACS~J1931.8-2635 &
0.352 &
$   501
^{+  441}_{ -221}$&
$  2114
^{+  355}_{ -311}$&
$  3.5
^{+  4.6}_{ -1.8}$&
$ 15.0
^{+  8.9}_{ -5.7}$&
$  6.7
\pm 0.40$
\\
RX~J1532.9+3021   &
0.363 &
$   293
^{+  433}_{ -114}$&
$  1544
^{+  191}_{ -210}$&
$  1.2
^{+  1.6}_{ -0.5}$&
$  5.9
^{+  2.5}_{ -2.1}$&
$  5.5
\pm 0.40$
\\
MACS~J1720.3+3536 &
0.391 &
$   505
^{+  248}_{ -162}$&
$  2055
^{+  204}_{ -204}$&
$  3.4
^{+  2.3}_{ -1.4}$&
$ 14.4
^{+  4.7}_{ -3.9}$&
$  6.6
\pm 0.40$
\\
MACS~J0416.1-2403 &
0.396 &
$   642
^{+  201}_{ -156}$&
$  1860
^{+  146}_{ -154}$&
$  3.4
^{+  1.5}_{ -1.1}$&
$ 10.7
^{+  2.7}_{ -2.4}$&
$  7.5
\pm 0.80$
\\
MACS~J0429.6-0253 &
0.399 &
$   394
^{+  238}_{ -143}$&
$  1792
^{+  225}_{ -208}$&
$  2.1
^{+  1.8}_{ -0.9}$&
$  9.6
^{+  4.1}_{ -3.0}$&
$  6.0
\pm 0.44$
\\
MACS~J1206.2-0847 &
0.440 &
$   587
^{+  248}_{ -176}$&
$  2181
^{+  165}_{ -178}$&
$  4.6
^{+  2.4}_{ -1.7}$&
$ 18.1
^{+  4.4}_{ -4.1}$&
$ 10.8
\pm 0.60$
\\
MACS~J0329.7-0211 &
0.450 &
$   254
^{+   95}_{  -63}$&
$  1697
^{+  129}_{ -127}$&
$  1.4
^{+  0.6}_{ -0.4}$&
$  8.6
^{+  2.1}_{ -1.8}$&
$  8.0
\pm 0.50$
\\
RX~J1347.5-1145   &
0.451 &
$   840
^{+  339}_{ -239}$&
$  2684
^{+  226}_{ -230}$&
$  9.8
^{+  5.6}_{ -3.6}$&
$ 34.2
^{+  9.4}_{ -8.1}$&
$ 15.5
\pm 0.60$
\\
MACS~J1149.5+2223 &
0.544 &
$  1108
^{+  404}_{ -291}$&
$  2334
^{+  169}_{ -178}$&
$ 10.8
^{+  5.4}_{ -3.7}$&
$ 25.0
^{+  5.8}_{ -5.3}$&
$  8.7
\pm 0.90$
\\
MACS~J0717.5+3745 &
0.548 &
$  1300
^{+  347}_{ -271}$&
$  2387
^{+  154}_{ -165}$&
$ 13.2
^{+  5.3}_{ -3.9}$&
$ 26.8
^{+  5.6}_{ -5.2}$&
$ 12.5
\pm 0.70$
\\
MACS~J0647.7+7015 &
0.584 &
$   468
^{+  254}_{ -160}$&
$  1884
^{+  189}_{ -192}$&
$  3.3
^{+  2.3}_{ -1.3}$&
$ 13.7
^{+  4.6}_{ -3.8}$&
$ 13.3
\pm 1.80$
\\
MACS~J0744.9+3927 &
0.686 &
$   574
^{+  269}_{ -192}$&
$  1982
^{+  179}_{ -185}$&
$  4.9
^{+  3.1}_{ -2.0}$&
$ 17.9
^{+  5.3}_{ -4.6}$&
$  8.9
\pm 0.80$
\\